\documentclass[aps,twocolumn,prd,showpacs,showkeys,preprintnumbers,nofootinbib,superscriptaddress,nobibnotes,floatfix,longbibliography]{revtex4-2}

\pdfoutput=1

\usepackage{amsmath}
\usepackage{amsfonts}
\usepackage{amssymb}
\usepackage{mathrsfs}
\usepackage{graphicx}
\usepackage{subfigure} 
\usepackage{color}
\usepackage[dvipsnames]{xcolor}
\usepackage{longtable}
\usepackage{bm}
\usepackage{blindtext}
\usepackage{wasysym}
\usepackage{hyperref}
\hypersetup{colorlinks=true,allcolors=blue}
\usepackage[normalem]{ulem}
\usepackage{lipsum}
\usepackage{cancel} 

\begin{document}
	

\title{Estimating the sensitivity of the IceCube Upgrade to probe the interior of the Earth using atmospheric neutrino oscillations}

\affiliation{III. Physikalisches Institut, RWTH Aachen University, D-52056 Aachen, Germany}
\affiliation{Department of Physics, University of Adelaide, Adelaide, 5005, Australia}
\affiliation{Dept. of Physics and Astronomy, University of Alaska Anchorage, 3211 Providence Dr., Anchorage, AK 99508, USA}
\affiliation{School of Physics and Center for Relativistic Astrophysics, Georgia Institute of Technology, Atlanta, GA 30332, USA}
\affiliation{Dept. of Physics, Southern University, Baton Rouge, LA 70813, USA}
\affiliation{Dept. of Physics, University of California, Berkeley, CA 94720, USA}
\affiliation{Lawrence Berkeley National Laboratory, Berkeley, CA 94720, USA}
\affiliation{Institut f{\"u}r Physik, Humboldt-Universit{\"a}t zu Berlin, D-12489 Berlin, Germany}
\affiliation{Fakult{\"a}t f{\"u}r Physik {\&} Astronomie, Ruhr-Universit{\"a}t Bochum, D-44780 Bochum, Germany}
\affiliation{Universit{\'e} Libre de Bruxelles, Science Faculty CP230, B-1050 Brussels, Belgium}
\affiliation{Vrije Universiteit Brussel (VUB), Dienst ELEM, B-1050 Brussels, Belgium}
\affiliation{Dept. of Physics, Simon Fraser University, Burnaby, BC V5A 1S6, Canada}
\affiliation{Department of Physics and Laboratory for Particle Physics and Cosmology, Harvard University, Cambridge, MA 02138, USA}
\affiliation{Dept. of Physics, Massachusetts Institute of Technology, Cambridge, MA 02139, USA}
\affiliation{Dept. of Physics and The International Center for Hadron Astrophysics, Chiba University, Chiba 263-8522, Japan}
\affiliation{Department of Physics, Loyola University Chicago, Chicago, IL 60660, USA}
\affiliation{Dept. of Physics and Astronomy, University of Canterbury, Private Bag 4800, Christchurch, New Zealand}
\affiliation{Dept. of Physics, University of Maryland, College Park, MD 20742, USA}
\affiliation{Dept. of Astronomy, Ohio State University, Columbus, OH 43210, USA}
\affiliation{Dept. of Physics and Center for Cosmology and Astro-Particle Physics, Ohio State University, Columbus, OH 43210, USA}
\affiliation{Niels Bohr Institute, University of Copenhagen, DK-2100 Copenhagen, Denmark}
\affiliation{Dept. of Physics, TU Dortmund University, D-44221 Dortmund, Germany}
\affiliation{Dept. of Physics and Astronomy, Michigan State University, East Lansing, MI 48824, USA}
\affiliation{Dept. of Physics, University of Alberta, Edmonton, Alberta, T6G 2E1, Canada}
\affiliation{Erlangen Centre for Astroparticle Physics, Friedrich-Alexander-Universit{\"a}t Erlangen-N{\"u}rnberg, D-91058 Erlangen, Germany}
\affiliation{Physik-department, Technische Universit{\"a}t M{\"u}nchen, D-85748 Garching, Germany}
\affiliation{D{\'e}partement de physique nucl{\'e}aire et corpusculaire, Universit{\'e} de Gen{\`e}ve, CH-1211 Gen{\`e}ve, Switzerland}
\affiliation{Dept. of Physics and Astronomy, University of Gent, B-9000 Gent, Belgium}
\affiliation{Dept. of Physics and Astronomy, University of California, Irvine, CA 92697, USA}
\affiliation{Karlsruhe Institute of Technology, Institute for Astroparticle Physics, D-76021 Karlsruhe, Germany}
\affiliation{Karlsruhe Institute of Technology, Institute of Experimental Particle Physics, D-76021 Karlsruhe, Germany}
\affiliation{Dept. of Physics, Engineering Physics, and Astronomy, Queen's University, Kingston, ON K7L 3N6, Canada}
\affiliation{Department of Physics {\&} Astronomy, University of Nevada, Las Vegas, NV 89154, USA}
\affiliation{Nevada Center for Astrophysics, University of Nevada, Las Vegas, NV 89154, USA}
\affiliation{Dept. of Physics and Astronomy, University of Kansas, Lawrence, KS 66045, USA}
\affiliation{UCLouvain, Centre for Cosmology, Particle Physics and Phenomenology, CP3, Chemin du Cyclotron 2, 1348 Louvain-la-Neuve, Belgium}
\affiliation{Department of Physics, Mercer University, Macon, GA 31207-0001, USA}
\affiliation{Dept. of Astronomy, University of Wisconsin{\textemdash}Madison, Madison, WI 53706, USA}
\affiliation{Dept. of Physics and Wisconsin IceCube Particle Astrophysics Center, University of Wisconsin{\textemdash}Madison, Madison, WI 53706, USA}
\affiliation{Institute of Physics, University of Mainz, Staudinger Weg 7, D-55099 Mainz, Germany}
\affiliation{Department of Physics, Marquette University, Milwaukee, WI 53201, USA}
\affiliation{Institut f{\"u}r Kernphysik, Universit{\"a}t M{\"u}nster, D-48149 M{\"u}nster, Germany}
\affiliation{Bartol Research Institute and Dept. of Physics and Astronomy, University of Delaware, Newark, DE 19716, USA}
\affiliation{Dept. of Physics, Yale University, New Haven, CT 06520, USA}
\affiliation{Columbia Astrophysics and Nevis Laboratories, Columbia University, New York, NY 10027, USA}
\affiliation{Dept. of Physics, University of Oxford, Parks Road, Oxford OX1 3PU, United Kingdom}
\affiliation{Dipartimento di Fisica e Astronomia Galileo Galilei, Universit{\`a} Degli Studi di Padova, I-35122 Padova PD, Italy}
\affiliation{Dept. of Physics, Drexel University, 3141 Chestnut Street, Philadelphia, PA 19104, USA}
\affiliation{Physics Department, South Dakota School of Mines and Technology, Rapid City, SD 57701, USA}
\affiliation{Dept. of Physics, University of Wisconsin, River Falls, WI 54022, USA}
\affiliation{Dept. of Physics and Astronomy, University of Rochester, Rochester, NY 14627, USA}
\affiliation{Department of Physics and Astronomy, University of Utah, Salt Lake City, UT 84112, USA}
\affiliation{Dept. of Physics, Chung-Ang University, Seoul 06974, Republic of Korea}
\affiliation{Oskar Klein Centre and Dept. of Physics, Stockholm University, SE-10691 Stockholm, Sweden}
\affiliation{Dept. of Physics and Astronomy, Stony Brook University, Stony Brook, NY 11794-3800, USA}
\affiliation{Dept. of Physics, Sungkyunkwan University, Suwon 16419, Republic of Korea}
\affiliation{Institute of Physics, Academia Sinica, Taipei, 11529, Taiwan}
\affiliation{Dept. of Physics and Astronomy, University of Alabama, Tuscaloosa, AL 35487, USA}
\affiliation{Dept. of Astronomy and Astrophysics, Pennsylvania State University, University Park, PA 16802, USA}
\affiliation{Dept. of Physics, Pennsylvania State University, University Park, PA 16802, USA}
\affiliation{Dept. of Physics and Astronomy, Uppsala University, Box 516, SE-75120 Uppsala, Sweden}
\affiliation{Dept. of Physics, University of Wuppertal, D-42119 Wuppertal, Germany}
\affiliation{Deutsches Elektronen-Synchrotron DESY, Platanenallee 6, D-15738 Zeuthen, Germany}

\author{R. Abbasi}
\affiliation{Department of Physics, Loyola University Chicago, Chicago, IL 60660, USA}
\author{M. Ackermann}
\affiliation{Deutsches Elektronen-Synchrotron DESY, Platanenallee 6, D-15738 Zeuthen, Germany}
\author{J. Adams}
\affiliation{Dept. of Physics and Astronomy, University of Canterbury, Private Bag 4800, Christchurch, New Zealand}
\author{S. K. Agarwalla}
\thanks{also at Institute of Physics, Sachivalaya Marg, Sainik School Post, Bhubaneswar 751005, India, and Homi Bhabha National Institute, Training School Complex, Anushakti Nagar, Mumbai 400094, India}
\affiliation{Dept. of Physics and Wisconsin IceCube Particle Astrophysics Center, University of Wisconsin{\textemdash}Madison, Madison, WI 53706, USA}
\author{J. A. Aguilar}
\affiliation{Universit{\'e} Libre de Bruxelles, Science Faculty CP230, B-1050 Brussels, Belgium}
\author{M. Ahlers}
\affiliation{Niels Bohr Institute, University of Copenhagen, DK-2100 Copenhagen, Denmark}
\author{J.M. Alameddine}
\affiliation{Dept. of Physics, TU Dortmund University, D-44221 Dortmund, Germany}
\author{S. Ali}
\affiliation{Dept. of Physics and Astronomy, University of Kansas, Lawrence, KS 66045, USA}
\author{N. M. Amin}
\affiliation{Bartol Research Institute and Dept. of Physics and Astronomy, University of Delaware, Newark, DE 19716, USA}
\author{K. Andeen}
\affiliation{Department of Physics, Marquette University, Milwaukee, WI 53201, USA}
% ...

\collaboration{IceCube Collaboration}
\email{analysis@icecube.wisc.edu}
\noaffiliation

\date{\today}

\begin{abstract}
    The IceCube Upgrade is a densely instrumented central region of the IceCube Neutrino Observatory, deployed during the 2025-26 polar season. It will reduce the detector's energy threshold and improve overall reconstruction capabilities for multi-GeV atmospheric neutrinos, which in turn enhance their sensitivity to Earth matter effects as they traverse through the deep Earth. In this study, we describe the potential of the IceCube Upgrade to observe Earth matter effects on atmospheric neutrinos and estimate the detector's sensitivity to probe key features of the Preliminary Reference Earth Model by utilizing these observations. We highlight the IceCube Upgrade's capability to estimate the mass of the Earth and verify the non-homogeneous distribution of matter density within the Earth. We also estimate the IceCube Upgrade sensitivity to measure the correlated densities of the Earth layers while incorporating constraints from the mass and moment of inertia of the Earth. Neutrino-based results would be independent and complementary to the seismic and gravitational measurements.
\end{abstract}


\maketitle

\section{Introduction and motivation}
\label{sec:introduction}

Understanding the structure of the deep interior of the Earth remains one of the most complex and intriguing challenges in geoscience. Despite significant technological advancements, direct observations beyond a few kilometers beneath the surface of the Earth are not feasible due to extreme environmental conditions like high temperature and pressure. As a result, indirect methods such as seismic studies~\cite{robertson1966interior,moser1982inner,loper1995core,alfe2007temperature,stacey2008physics,McDonough:2022,Thorne:2022,Hirose:2022} and gravitational measurements~\cite{Ries:1992,Williams:1994,Luzum:2011,Rosi:2014kva,astro_almanac,Chen:2014} have been employed to probe the internal structure of the Earth. The information obtained from seismic wave studies indicates that the Earth has a layered structure composed of concentric spherical shells. These layers are broadly classified into the mantle and the core, each containing multiple sub-layers, with the outermost layer, the crust, being relatively thin in comparison. As we move from the surface of the Earth toward the center, the densities of these layers increase. This increment is not always gradual, but could feature sharp density jumps at layer boundaries. Using seismic wave velocity data, several Earth density models have been developed over the past few decades~\cite{Gilbert-Dziewonski1975,Dziewonski-Hales-Lapwood1975,Kennett-Engdahl1991,Kennett-etal1995,Cammarano-etal-2005,Kustowski-etal-2008,Xiaolong2021}. Among these, the Preliminary Reference Earth Model (PREM)~\cite{Dziewonski:1981xy} is the most widely used one-dimensional radial density model.

Although extensive seismic and gravitational data have significantly enhanced our understanding of the Earth's interior, many open questions remain unresolved~\cite{McDonough:2022}. For example, the mass and the chemical composition of the core, as well as the precise density transition at the inner and outer core boundary, are still not well known~\cite{McDonough:2024}. Similarly, the amount of light elements, such as hydrogen in the core and the amount of water present in the mantle, remain uncertain~\cite{williams2001hydrogen,Hirose:2021,Hirose:2022}. Moreover, the mantle’s density is uncertain by about 5\%, while the core remains the most uncertain region inside the Earth~\cite{Bolt:1991,kennett:1998,Masters:2003}. Achieving a more precise understanding of these unknowns is crucial for elucidating dynamic processes, thermal structure, and geochemical interactions of the Earth that shape its long-term evolution. Therefore, complementary and independent probes could further improve our understanding of the internal structure of the Earth.

In addition to the above-mentioned traditional methods, atmospheric neutrinos provide a unique and complementary probe of the interior of the Earth. These neutrinos are produced in the interactions of high-energy cosmic rays with the atmosphere of the Earth. Their flux consists of neutrinos and antineutrinos of electron and muon flavors. Atmospheric neutrinos span a broad energy range, from a few MeV to above TeV, and traverse a wide range of baselines, from tens of kilometers to about 13,000 km, depending on their arrival direction with respect to the detector position. At energies above 10 TeV, the increase in neutrino-nucleon cross section~\cite{Gandhi:1995tf,IceCube:2017roe} leads to significant attenuation of the high-energy atmospheric neutrino flux during propagation through the Earth. The resulting attenuation depends on the average nucleon number density along the neutrino trajectory. Therefore, by measuring the energy- and zenith-dependent suppression of the neutrino flux, we can infer the Earth's radial density distribution. On the other hand, in the multi-GeV energy range, neutrinos undergo flavor oscillations that are sensitive to the electron number density distribution along the neutrino trajectory within the Earth. Therefore, measurements of matter-induced neutrino oscillations provide a tool to probe the internal structure of the Earth.

The idea of using the attenuation of high-energy neutrinos to learn about the interior of Earth was first proposed in Refs.~\cite{Placci:1973,Volkova:1974xa}. Subsequently, several studies~\cite{Gonzalez-Garcia:2007wfs,Donini:2018tsg,IceCube:2025utw} demonstrated the feasibility of neutrino absorption tomography as a method for probing the Earth's internal density structure. A recent study in Ref.~\cite{IceCube:2025utw} probed the Earth’s density profile using approximately 11 years of high-energy atmospheric neutrino data observed by the IceCube detector in the energy range from 500 GeV - 100 TeV. Compared to a notable study in Ref.~\cite{Donini:2018tsg}, which used one year of IceCube data in the TeV energy range, Ref.~\cite{IceCube:2025utw} provides the most precise determination of the Earth’s density profile to date using weak interactions of neutrinos. This improvement is primarily driven by the large statistics, improved energy reconstruction capabilities, and a more updated treatment of systematic uncertainties. 

\begin{figure*}[htp!]
	\includegraphics[width=\linewidth]{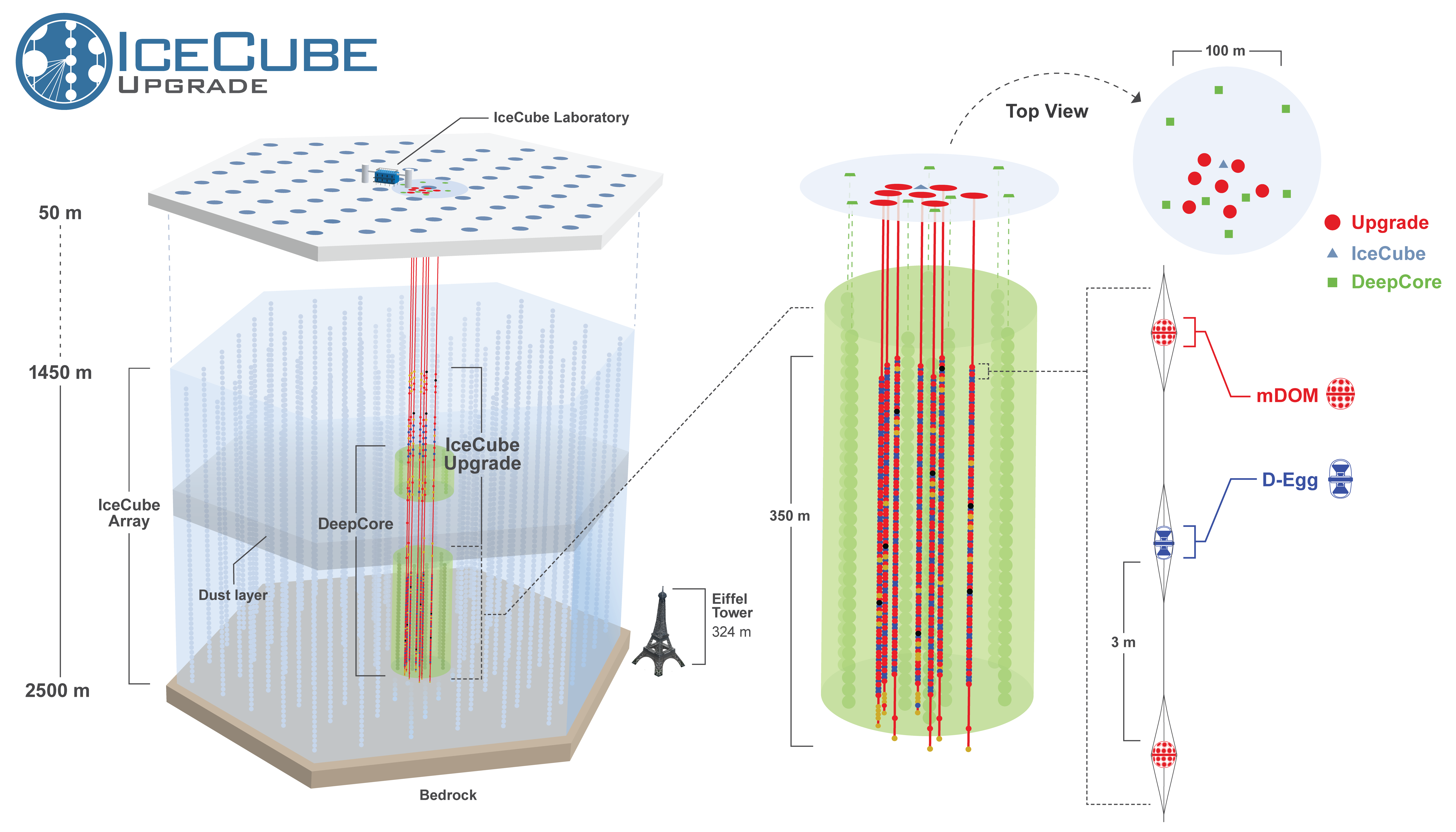}
	\caption{Schematic overview of the IceCube Upgrade in its originally planned configuration of seven additional strings within the existing IceCube Neutrino Observatory. It illustrates the spatial layout and instrumentation of the main IceCube array (light blue), the DeepCore subarray (green), and the IceCube Upgrade region (highlighted in the center column). The inset on the right zooms into the IceCube Upgrade region, which includes seven additional densely instrumented strings, shown in red, and their positions relative to surrounding IceCube (blue triangles) and DeepCore (green squares) strings. The Upgrade strings are equipped with two new types of optical sensors: the multi-PMT DOM (mDOM) shown in red, and the dual-PMT D-Egg module shown in blue.}
	\label{fig:upgrade_schematic}
\end{figure*}

As far as atmospheric neutrinos with GeV energies are concerned, they undergo flavor oscillations while propagating through the Earth. In addition to flavor oscillations, they  also undergo charged-current (CC) coherent forward scattering with ambient electrons, which gives rise to an effective matter potential. This potential modifies the neutrino oscillation patterns relative to the vacuum oscillation scenario. This phenomenon is known as matter effects. The dependence of these matter effects on the electron density distribution inside the Earth can be used to perform neutrino oscillation tomography~\cite{Nicolaidis:1990jm}. Unlike the aforementioned methods, this approach probes the internal electron density profile of the Earth, which may also provide constraints on its chemical composition. Several sensitivity studies~\cite{IceCube-PINGU:2014okk,Winter:2015zwx,Rott:2015kwa,Bourret:2020zwg,Kumar:2021faw,Denton:2021rgt,Capozzi:2021hkl,Kelly:2021jfs,Upadhyay:2021kzf,Maderer:2022toi,DOlivoSaez:2022vdl,Upadhyay:2022jfd,Raikwal:2023jkf,Jesus-Valls:2024tgd,Upadhyay:2024gra,Chattopadhyay:2025cgt,Upadhyay:2026kfq,Krishnamoorthi:2026yns,Chattopadhyay:2026eev} have demonstrated that neutrino oscillations can be used to probe the information about the internal structure and chemical composition of the Earth. The sensitivity study presented in this work exploits matter-induced modifications in neutrino oscillation patterns using atmospheric neutrinos with the IceCube Upgrade detector.

The IceCube Neutrino Observatory~\cite{IceCube:2016zyt} is a cubic-kilometer detector located at the South Pole, designed to detect high-energy neutrinos in the TeV--PeV energy range. It consists of 5,160 photodetectors, known as Digital Optical Modules (DOMs), deployed deep within the Antarctic ice on 86 vertical cables, called strings, from a depth of 1.45 km to 2.45 km. IceCube has a denser sub-array in the central region, known as IceCube DeepCore~\cite{IceCube:2011ucd}, to enhance sensitivity to lower-energy neutrinos, especially in the multi-GeV range. DeepCore includes eight additional strings with closer vertical spacing between DOMs, which are deployed at depths from 2.1 km to 2.4 km. DeepCore has accumulated more than ten years of atmospheric neutrino data, with its most recent neutrino oscillation results presented in Refs.~\cite{IceCubeCollaboration:2024ssx,IceCubeCollaboration:2023wtb}. The current detector configuration, including both IceCube and DeepCore strings, is collectively referred to as ``IC86'' in this work.

The recent extension of the IceCube detector is known as the IceCube Upgrade. Figure~\ref{fig:upgrade_schematic} shows the originally proposed configuration of the IceCube Upgrade. This extension, recently deployed during the austral summer season of 2025-2026, added six new strings instead of the originally planned seven strings within the DeepCore fiducial volume. However, out of these six strings, only five are operational as one string is unresponsive. These five new strings bring the total number of working strings in IceCube to 91 (referred to as the ``IC91'' configuration)  instead of 93 (referred to as the ``IC93'' configuration). Note that even though IC91 will be operating, we have considered IC93 configuration in the present work. These new strings are equipped with advanced optical sensor modules, including the ``multi-PMT Digital Optical Modules'' (mDOMs)~\cite{IceCube:2019anq, IceCube:2021eij,Mechbal:2023fic} and the ``Dual optical sensors in an Ellipsoid Glass for Gen2'' (D-Eggs)~\cite{IceCube:2022mng}. In addition to these sensors, the Upgrade also includes new calibration devices~\cite{Henningsen:2020zsj,Khera:2021npv,Rongen:2021rgc,IceCube:2021zse,Kang:2023kjl,Anthony:2023srl} to better understand the optical properties of the Antarctic ice, and hence, improve event reconstruction, particularly for low-energy events. With these improvements, and the denser instrument spacing, the IceCube Upgrade will lower the energy threshold and significantly enhance the atmospheric neutrino detection rate in the GeV range. Recent work in Ref.~\cite{IceCube:2025chb} presents the sensitivity studies using atmospheric neutrino oscillations with the IceCube Upgrade considering the IC93 configuration. 

The present study utilizes simulated neutrino oscillation data to explore the sensitivity of the IceCube Upgrade for probing key features of the PREM density profile. To achieve it, this work considers an Earth density model composed of 12 concentric spherical shells, each with a constant density equal to the volume-averaged density of the fine layers of PREM within that shell. This ensures that the mass of each shell, and consequently the total mass of the Earth, is the same as in the PREM model. This approximated model is referred to as the ``12-layered PREM profile''. This work discusses the expected sensitivity of the IceCube Upgrade to establish the presence of Earth matter effects in atmospheric neutrino oscillations by rejecting the vacuum oscillation scenario. Utilizing these matter effects, we further evaluate the ability of the detector to distinguish a layered density structure from a homogeneous Earth profile. The study further investigates the potential of the IceCube Upgrade to measure the total mass of the Earth and to constrain the correlated densities of individual layers inside the Earth. The sensitivity results are also compared to those expected from IceCube DeepCore, assuming the same exposure, to demonstrate the improvements offered by the IceCube Upgrade extension. 

These studies are also performed using 9.3 years of data from IceCube DeepCore. Additional information about this data sample can be found in Ref.~\cite{IceCubeCollaboration:2024ssx}. The preliminary sensitivity results for these studies using Monte Carlo (MC) equivalent to 9.3 years of data sample are presented in Ref.~\cite{Chattopadhyay:2025cgt,Upadhyay:2026kfq,Krishnamoorthi:2026yns,Chattopadhyay:2026eev}.

The description of Earth matter effects is outlined in Section~\ref{sec:matter_effects}, followed by the event selection used for the IceCube Upgrade in Section~\ref{sec:events_at_icecube}. The details of the analysis procedure, along with the systematic uncertainty of the parameters considered, are discussed in Section~\ref{sec:analysis_methodology}. Section~\ref{sec:sensitivity_result} presents the sensitivity results, with a discussion on the impact of systematic uncertainties. Finally, Section~\ref{sec:conclusion} summarizes the conclusions of this study. Appendix~\ref{app:prob} discusses the effects of various Earth density profiles on neutrino oscillograms.

\section{Earth matter effects in neutrino oscillations}
\label{sec:matter_effects}

Neutrino oscillations arising from mixing between neutrino flavors and mass eigenstates are described by the Pontecorvo-Maki-Nakagawa-Sakata (PMNS) matrix~\cite{Pontecorvo:1957cp,Maki:1962mu}. The oscillation probabilities are characterized by three mixing angles ($\theta_{12}$, $\theta_{13}$, and $\theta_{23}$), a Dirac CP-violating phase ($\delta_{\rm CP}$), and two mass-squared differences ($\Delta m^2_{21}$ and $\Delta m^2_{31}$). These oscillation parameters are measured with high precision~\cite{Capozzi:2025wyn,Esteban:2024eli,NuFIT,deSalas:2020pgw}. In addition to oscillation parameters, neutrino oscillation probabilities also depend on the medium in which neutrinos propagate. As atmospheric neutrinos travel through the Earth, they experience an effective charged-current matter potential, which is given by:
\begin{equation}
	V_\text{CC} = \pm \sqrt{2} G_F N_e \approx \pm \, 7.6 \times Y_e \times 10^{-14} \left[\frac{\rho}{\text{g/cm}^3}\right]~\text{eV}\,,
	\label{eq:vcc}
\end{equation}
where $G_F$ is the Fermi coupling constant and $N_e$ is the electron number density inside matter with density $\rho$. The $\pm$ sign corresponds to neutrinos and antineutrinos, respectively. Here, $Y_e$ denotes the electron-to-nucleon number density ratio inside the Earth and is defined as $Y_e = N_e/(N_p + N_n)$, where $N_p$ and $N_n$ represent the proton and neutron number densities, respectively. It is clear from Eq.~\ref{eq:vcc} that the neutrino oscillations are sensitive to the product $\rho \times Y_e$. One can choose to constrain either the matter density ($\rho$) or the chemical composition ($Y_e$) of the Earth, assuming the other is perfectly known. As shown in recent studies~\cite{Rott:2015kwa,Maderer:2022toi}, neutrino data are not yet sufficiently sensitive to measure $Y_e$ more precisely than their present allowed range. Therefore, in the present work, our analyses focus on constraining different matter density profiles, assuming that the chemical composition of the Earth is precisely known. The following $Y_e$ values are considered for the 12-layered PREM profile: $Y_e = 0.4656$ for both the inner and outer core (corresponding to a pure FeNi core) and $Y_e = 0.4957$ for the mantle (corresponding to a pyrolite mantle) as proposed in Ref.~\cite{Rott:2015kwa}. For the case of a uniform matter density profile, $Y_e = 0.5$ is considered assuming the Earth to be neutral and isoscalar. 

\begin{figure*}[htp!]
	\includegraphics[width=\linewidth]{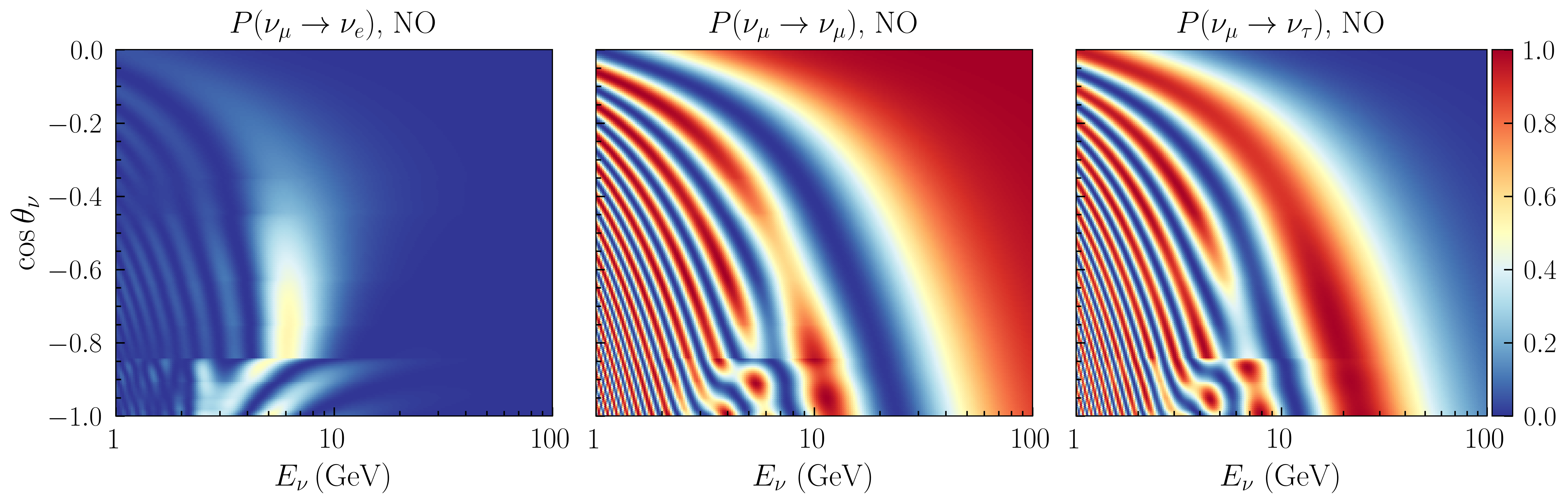}
	\caption{Three-flavor muon neutrino oscillation probabilities in the $(E_\nu, \cos\theta_\nu)$ plane: $P(\nu_\mu \rightarrow \nu_e)$ appearance probability (left panel), $P(\nu_\mu \rightarrow \nu_\mu)$ survival probability (middle panel), and $P(\nu_\mu \rightarrow \nu_\tau)$ appearance probability (right panel). These probabilities are computed using the 12-layered PREM density model of the Earth (black curve in Fig.~\ref{fig:radial_density_profile}), assuming normal mass ordering, with the nominal oscillation parameter values given in Table~\ref{tab:osc-param-value}.}
	\label{fig:prem_osc}
\end{figure*}

\begin{table}
	\centering
	\begin{tabular}{c @{\hskip 15pt} c}
		\hline
		\hline
		Oscillation parameters & Values \\
		\hline
		$\theta_{12}$ & 33.41$^\circ$ \\
		$\theta_{23}$ & 47.50$^\circ$ \\
		$\theta_{13}$ & 8.54$^\circ$ \\
		$\Delta m^2_{31}$ & $2.47\times 10^{-3}$ (eV$^2$) \\
		$\Delta m^2_{21}$ & $7.41\times10^{-5}$ (eV$^2$) \\
		$\delta_{\rm CP}$ & 0$^\circ$ \\
		Mass Ordering & Normal (NO)\\
		\hline
		\hline 
	\end{tabular}
	\caption{The nominal values of neutrino oscillation parameters considered in this work. These values are taken from NuFit 5.2~\cite{Esteban:2020cvm}, except for $\delta_{\rm CP}$, $\theta_{23}$, and $\Delta m^2_{31}$. The values of $\theta_{23}$ and $\Delta m^2_{31}$ are the best-fit values as reported by the IceCube DeepCore measurement~\cite{IceCubeCollaboration:2024ssx}. The value of $\delta_\text{CP}$ is taken to be zero, as it has negligible impact on the analyses performed in this work.}
	\label{tab:osc-param-value}
\end{table}

The matter potential effectively results in modified mixing angles and mass-squared splittings, thereby altering the oscillation probabilities of neutrinos as they pass through the Earth. In particular, the resonant enhancement of a small mixing angle, $\theta_{13}$, to an effective value as large as $45^\circ$ is known as the Mikheyev-Smirnov-Wolfenstein (MSW) resonance~\cite{Wolfenstein:1977ue, Mikheev:1986gs, Mikheev:1986wj}. The MSW resonance occurs for neutrinos in the case of normal mass ordering (NO) and for antineutrinos in the case of inverted mass ordering (IO). The occurrence of the MSW resonance depends on both the neutrino energy and the ambient electron density it encounters along its path. In particular, the MSW resonance is most prominent for neutrinos passing through the mantle with energies in the range of 6 to 10 GeV. In addition, neutrinos with core-passing trajectories experience a sharp density jump at the core-mantle boundary (CMB), which significantly modifies their oscillation probabilities, particularly for those with energies between 3 and 6 GeV. This occurs when the periodic changes in matter density align with the intrinsic oscillation phase, resulting in constructive interference. This phenomenon is known as the neutrino oscillation length resonance (NOLR)~\cite{Petcov:1998su, Chizhov:1998ug, Petcov:1998sg, Chizhov:1999az, Chizhov:1999he} or parametric resonance (PR)~\cite{Ermilova:1986, Akhmedov:1988kd, Krastev:1989, Akhmedov:1998ui, Akhmedov:1998xq}. These matter resonances are illustrated in Fig.~\ref{fig:prem_osc}, which shows the $P(\nu_\mu \rightarrow \nu_e)$ appearance probabilities (left panel), $P(\nu_\mu \rightarrow \nu_\mu)$ survival probabilities (middle panel), and $P(\nu_\mu \rightarrow \nu_\tau)$ appearance probabilities (right panel) as a function of true neutrino energy $(E_\nu)$ and arrival direction ($\cos\theta_\nu$). The probabilities are calculated using the three-flavor neutrino oscillations in the presence of Earth matter effects, assuming the 12-layered PREM density profile of the Earth (see Fig.~\ref{fig:radial_density_profile}). These probabilities are obtained using the nominal values of neutrino oscillation parameters given in Table~\ref{tab:osc-param-value}. The blue (red) diagonal band in the middle (right) panel, extending from ($E_\nu = 1$ GeV, $\cos\theta_\nu = 0$) to ($E_\nu = 30$ GeV, $\cos\theta_\nu = -\,1$), corresponds to the first vacuum oscillation minimum (maximum), also referred to as the ``oscillation valley''~\cite{Kumar:2020wgz,Kumar:2021lrn}. In each panel, distortions around $-\,0.8 < \cos\theta_\nu < -\,0.5$ and $6~\text{GeV} < E_\nu < 10~\text{GeV}$ are caused by the MSW resonance, while those around $\cos\theta_\nu < -\,0.8$ and $3~\text{GeV} < E_\nu < 6~\text{GeV}$ arise due to the NOLR/PR resonance.

\section{Event Simulation of the IceCube Upgrade}
\label{sec:events_at_icecube}

The IceCube Upgrade simulation chain closely follows the procedure used for IceCube DeepCore simulations in previous studies and is described in details in Ref.~\cite{IceCubeCollaboration:2023wtb}. The simulated Monte Carlo (MC) dataset includes neutrinos of all flavors, and backgrounds contributed by atmospheric muons and detector noise. In IceCube, the incoming neutrino interacts with ice and produces secondary charged particles which emit Cherenkov photons. The neutrino events are classified into track-like or cascade-like topologies based on the spatial and temporal distribution of Cherenkov light. Track-like signatures are typically associated with $\nu_{\mu}$ CC interactions, where the resulting muon travels a long distance through the detector, producing an elongated light pattern. In contrast, cascade-like topologies arise from $\nu_e$ and $\nu_\tau$ CC interactions, as well as from neutral-current (NC) interactions, which produce more spherical light patterns. However, some of the $\nu_\tau$ CC interactions could also produce $\tau$ leptons, which decay into muons resulting in track-like events.

A major task in the event selection is separating neutrino signals from background events. To suppress backgrounds and obtain a neutrino-dominated sample, a series of filtering stages is applied. The noise-suppressed sample is then processed using the DynEdge-based reconstruction algorithm~\cite{Abbasi:2022ypr}. Two DynEdge models are trained separately to reconstruct neutrino energy and direction ($\cos\theta_\nu$). A third DynEdge model is trained to predict particle identification (PID) scores classifying events into track-like and cascade-like topologies. Lastly, a fourth DynEdge model separates neutrinos from atmospheric muons. The details of the filtering levels, the detector response, event simulation, event selection, and reconstruction method are provided in Ref.~\cite{IceCube:2025chb}.

\begin{figure*}[htp!]
	\includegraphics[width=\linewidth]{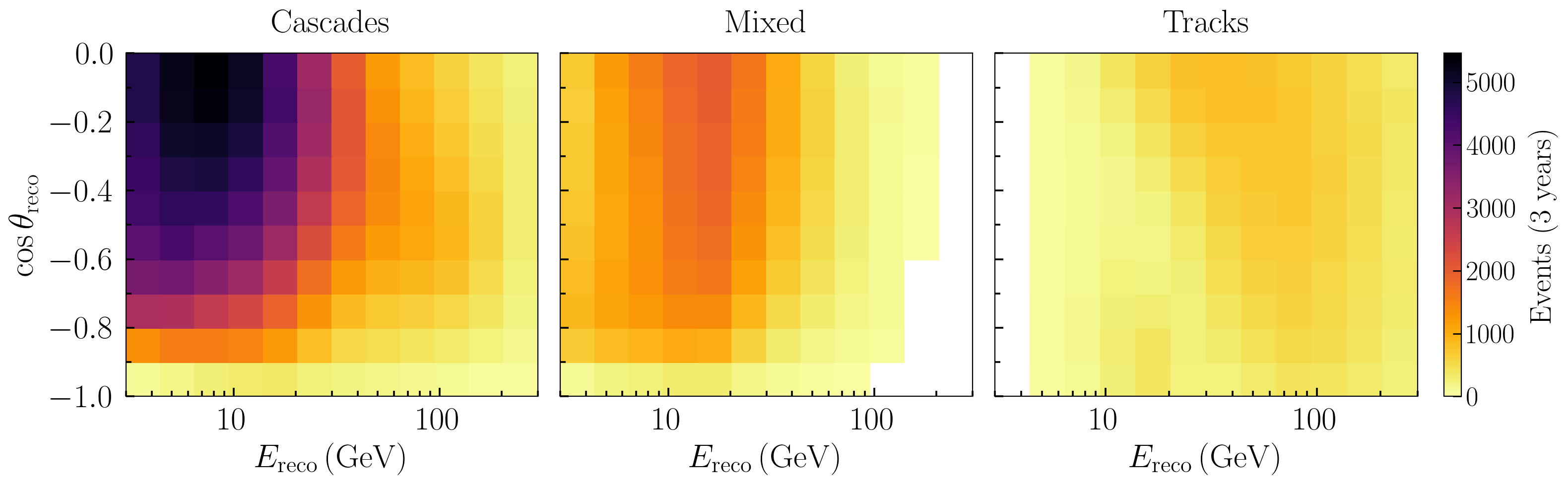}
	\caption{Distribution of expected events in the analysis binning scheme for 3 years exposure using the IC93 detector configuration, assuming the 12-layered PREM density profile (black curve in Fig.~\ref{fig:radial_density_profile}). Events are binned according to reconstructed neutrino energy ($E_\text{reco}$), reconstructed zenith angle ($\cos\theta_\text{reco}$), and particle identification category (cascade-like, mixed, and track-like).}
	\label{fig:event_dist}
\end{figure*}

In the present study, further cuts based on the reconstructed energy ($E_\text{reco}$) and reconstructed cosine zenith ($\cos\theta_\text{reco}$) are applied on the processed sample. The reconstructed energy and reconstructed cosine zenith are taken in the ranges of 3 GeV to 300 GeV and $-\,1$ to 0, respectively, which selects only upward-going events. To obtain the expected events, the MC events are weighted using the Honda atmospheric neutrino flux~\cite{Honda:2015fha} and the neutrino oscillation probabilities. The oscillation parameters used in the calculation of three-flavor oscillation probabilities are given in Table~\ref{tab:osc-param-value}. These expected events are binned into 12 logarithmically spaced energy bins and 10 linearly spaced cosine zenith bins. Based on the PID score, the neutrino sample is categorized into three classes: cascade-like, mixed, and track-like. The distribution of the expected events is shown in Fig.~\ref{fig:event_dist}. To reduce the impact of statistical fluctuations, bins with low event statistics are excluded from the analysis, which are left blank in Fig.~\ref{fig:event_dist}. These distributions are obtained using the nominal values of oscillation parameters and the nuisance parameters provided in Table~\ref{tab:systematic_params}.

These expected events are then used to study the performance of the IceCube Upgrade to the Earth matter effects. In addition to the IceCube Upgrade events, MC simulations of the IceCube DeepCore configuration, previously used in Ref.~\cite{IceCubeCollaboration:2024ssx}, are also included in this study for comparison and to perform combined sensitivity analyses. The simulation settings, event selection, reconstruction methods, and analysis-level cuts for the DeepCore dataset follow those described in Ref.~\cite{IceCubeCollaboration:2024ssx}. This IceCube DeepCore sample is binned into 10 logarithmically spaced energy bins between 5 and 100 GeV, 8 linearly spaced cosine zenith bins in the range [$-\,1$, 0.04], and 3 PID bins: cascades, mixed, and tracks.

\section{Analysis}
\label{sec:analysis_methodology}

\subsection{Analysis setup}
\label{sec:analysis_setup}

\begin{figure}[htp!]
	\includegraphics[width=\linewidth]{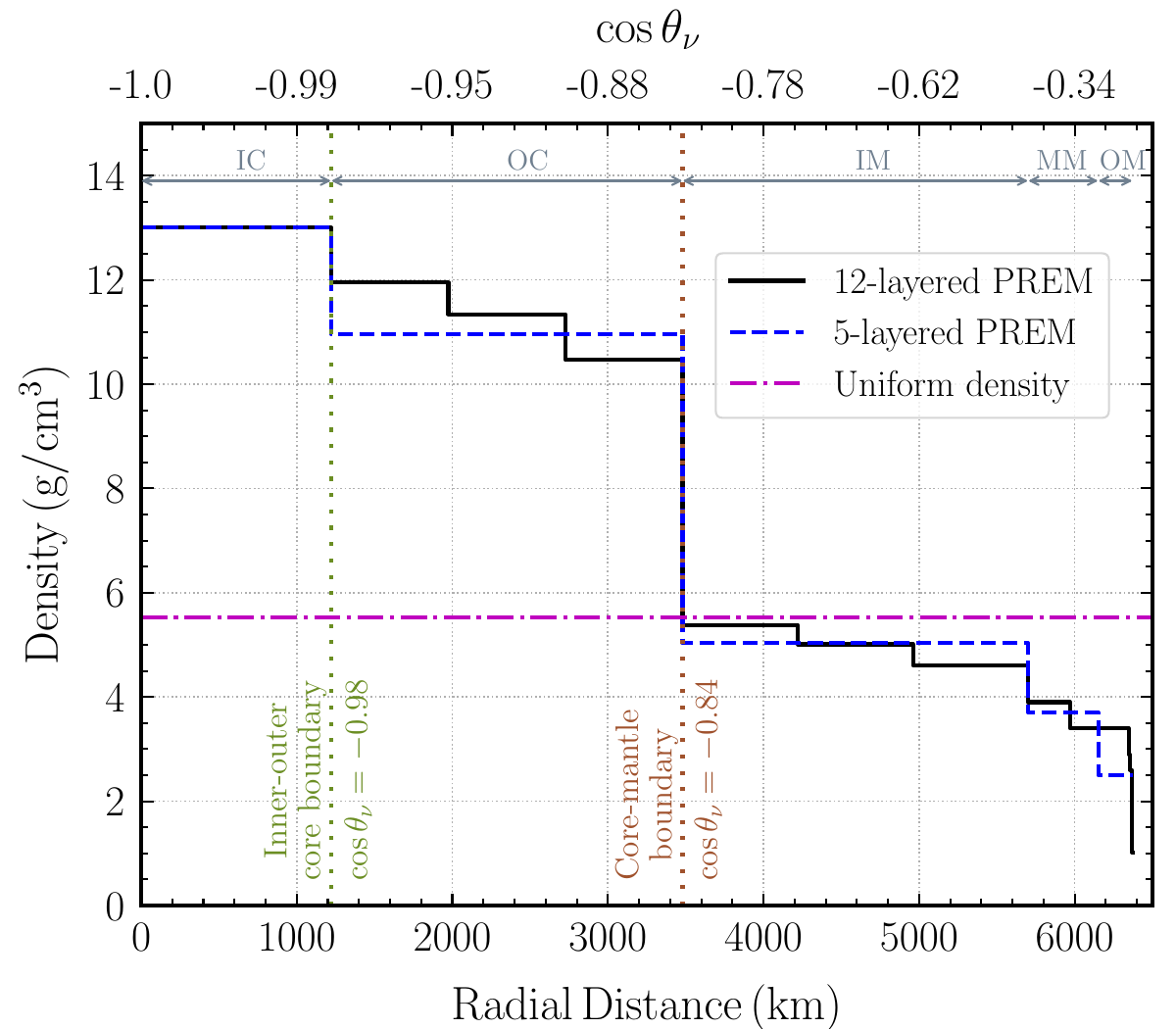}
	\caption{Density distribution as a function of radial distance from the center of the Earth. The black, dashed blue, and dot-dashed magenta curves correspond to the 12-layered PREM, 5-layered PREM, and homogeneous (uniform) Earth density models, respectively. The top x-axis shows the cosine of the zenith angle ($\cos\theta_\nu$) corresponding to neutrino trajectories, related to the radial distance $R$ (the shortest distance from the center of the Earth to the neutrino trajectory) by the relation $\cos\theta_\nu \approx \sqrt{1 - (R/R_E)^2}$, where $R_E$ is the radius of the Earth. The vertical dotted green and brown lines indicate the inner-outer core boundary and the core-mantle boundary, corresponding to $\cos\theta_\nu = - \, 0.98$ and $\cos\theta_\nu = - \, 0.84$, respectively.}
	\label{fig:radial_density_profile}
\end{figure}

This sensitivity study addresses four main objectives: establishing matter effects, validating the layered structure inside the Earth, and measuring the total mass and correlated layer densities of the Earth. To achieve the first three objectives, this work considers a 12-layered Earth density model as a true hypothesis. To establish Earth matter effect, we consider vacuum as a test hypothesis. To validate the layered structure, we take uniform density profile as another test hypothesis. For the analysis of the correlated density measurement, a simplified 5-layered model is adopted as the true hypothesis, where the five layers are the inner core (IC), outer core (OC), inner mantle (IM), middle mantle (MM), and outer mantle (OM). The 5-layered model has been obtained by merging the finer layers of the 12-layered PREM model, and the densities of the merged layers are calculated by taking the volume average of the densities of the finer layers. While doing this, we kept the layer boundaries of the 5-layered model the same as those in the 12-layered model. Note that, while deriving the 5-layered PREM model, we ensure that the total mass and moment of inertia of the Earth match those considered in the 12-layered PREM model. Figure~\ref{fig:radial_density_profile} illustrates these density models.

For the mass of the Earth measurement, the density of each layer in the 12-layered PREM profile is scaled uniformly using a single scaling factor ($\mathrm{\alpha}$), allowing variations in the total mass of the Earth without imposing external constraints from seismic or gravitational studies. For the correlated density measurement, the densities of the IC, OC, IM, and MM are simultaneously varied while satisfying the following constraints: (i) the total mass and moment of inertia of the Earth are kept fixed, as these quantities are extremely well known from gravitational measurements, (ii) the density of the OM is kept constant, as it has comparatively small uncertainties, and (iii) the ratio of the densities between the IC and OC is fixed to the value given in the 5-layered PREM profile for simplicity. While the core, inner mantle, and middle mantle have different scaling factors ($\mathrm{\alpha_C}$, $\mathrm{\alpha_{IM}}$, and $\mathrm{\alpha_{MM}}$, respectively), the imposition of the above-mentioned external constraints allows the analysis to be effectively parameterized in terms of any one of these scaling factors. Here, $\mathrm{\alpha_C}$ is taken as the physics parameter to be measured. This study provides sensitivities to estimate these scaling factors ($\mathrm{\alpha}$ and $\mathrm{\alpha_C}$) using simulated atmospheric neutrino data with the IceCube Upgrade.

It should be noted that at least five layers are required to impose the above-mentioned constraints, where one free scaling factor remains, which can be estimated in this study. These five layers are sufficient to estimate the correlated densities, as current neutrino measurements are primarily sensitive to large-scale density variations. The advantages of using a larger number of layers are limited by the energy and zenith angle resolution of the detector. Therefore, a 5-layered model is adopted in this analysis.

\begin{figure*}[htp!]
	\includegraphics[width=\linewidth]{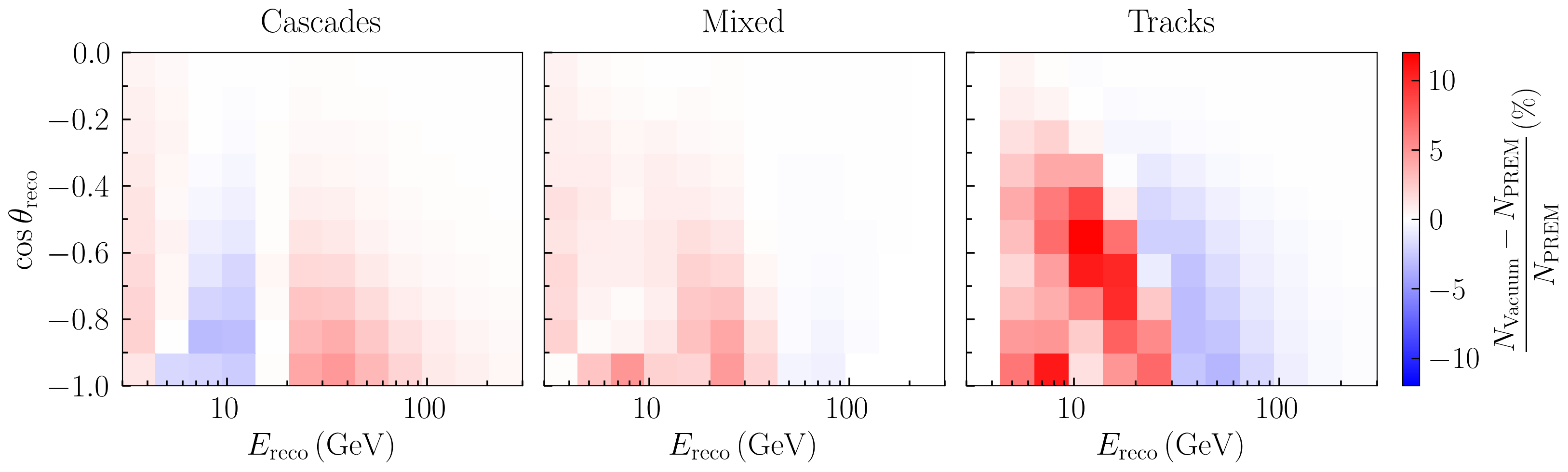}
	\caption{Relative difference in expected event counts between the vacuum and 12-layered PREM hypotheses for 3 years of IceCube Upgrade (IC93) exposure. Events are binned in reconstructed neutrino energy ($E_\text{reco}$), reconstructed zenith angle ($\cos\theta_\text{reco}$), and particle identification category (cascade-like, mixed, and track-like).}
	\label{fig:v_event_diff}
\end{figure*}

\begin{figure*}[htp!]
	\includegraphics[width=\linewidth]{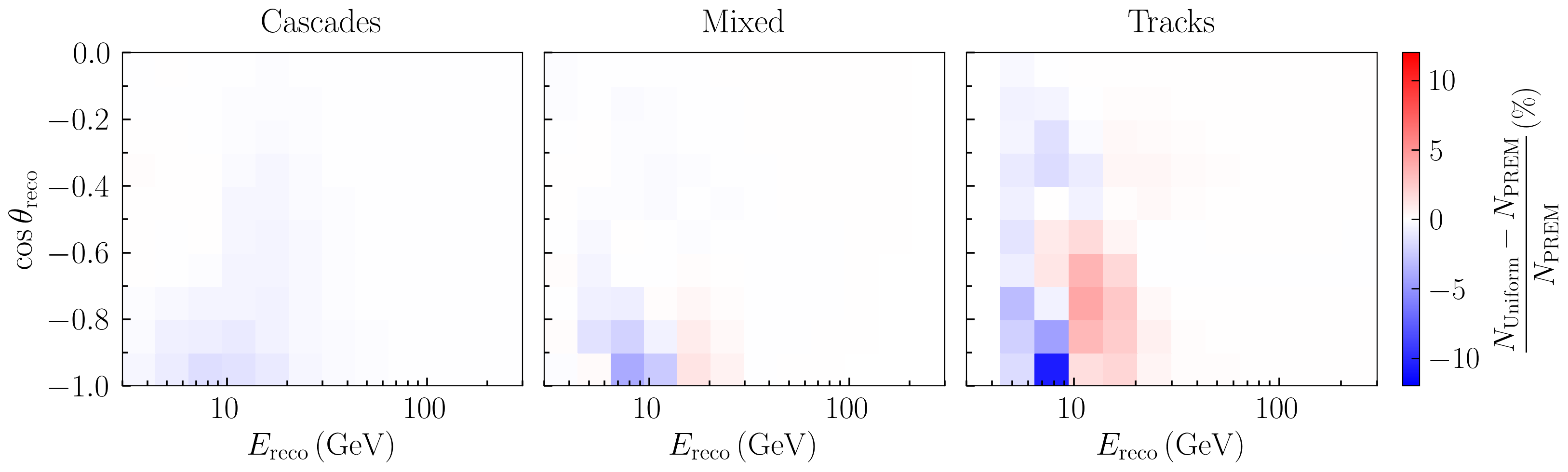}
	\caption{Relative difference in expected event counts between the uniform and 12-layered PREM hypotheses for 3 years of IceCube Upgrade (IC93) exposure. Events are binned in reconstructed neutrino energy ($E_\text{reco}$), reconstructed zenith angle ($\cos\theta_\text{reco}$), and particle identification category (cascade-like, mixed, and track-like).}
	\label{fig:u_event_diff}
\end{figure*}

\begin{figure*}[htp!]
	\includegraphics[width=\linewidth]{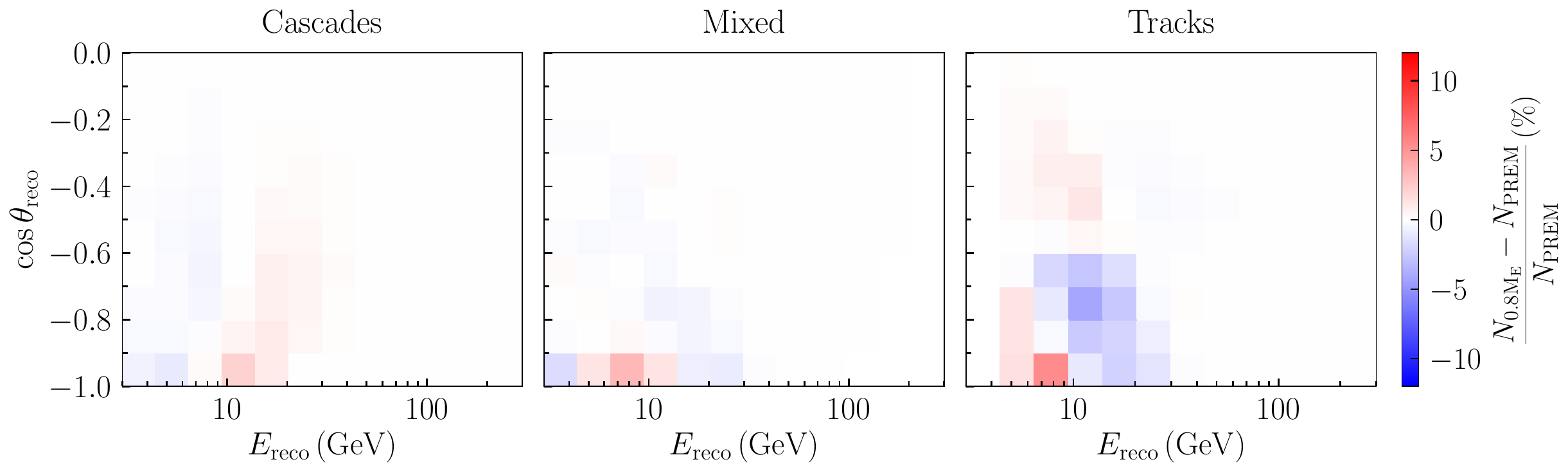}
	\caption{Relative difference in expected event counts between the 12-layered PREM profile and a modified Earth density profile with the total mass reduced by 20\% for 3 years of IceCube Upgrade (IC93) exposure. Events are binned in reconstructed neutrino energy ($E_\text{reco}$), reconstructed zenith angle ($\cos\theta_\text{reco}$), and particle identification category (cascade-like, mixed, and track-like).}
	\label{fig:m_event_diff}
\end{figure*}

In this study, the main signals arise from the effects of matter resonances induced in the neutrino oscillation probabilities under different Earth density profiles. A more detailed discussion is provided in the appendix~\ref{app:prob}. To identify the expected signal regions in terms of the reconstructed neutrino variables ($E_\text{reco}, \cos\theta_\text{reco}$), the relative difference in event counts are plotted for various Earth density models for the IC93 configuration in Figs.~\ref{fig:v_event_diff}, \ref{fig:u_event_diff}, and \ref{fig:m_event_diff}. These plots use the nominal values of oscillation parameters taken from Table~\ref{tab:osc-param-value}, and nuisance parameters from Table~\ref{tab:systematic_params}. For all three analyses, the signal regions arise where matter effects dominate, \textit{i.e.}, low-energy and high-baseline regions. However, for the analysis establishing  Earth matter effect, noticeable differences also extend into the regions of  higher energies and lower baselines.

An additional feature can be observed in Fig.~\ref{fig:v_event_diff}, where the relative event difference between the vacuum oscillation and the 12-layered PREM scenario is positive below $E_\text{reco} \approx 30$ GeV and negative above it. This behavior arises because the effective value of $\Delta m^2_{31}$ in the presence of matter effects differs from the vacuum oscillation case, leading to a shift in the oscillation valley. This shift leads to a transition from positive to negative (red to blue) in the relative event difference plot.

The analysis uses the simulated MC events generated without any statistical fluctuations to estimate the median sensitivity. This MC template is also referred to as the ``Asimov dataset''~\cite{Cowan:2010js}. A modified $\chi^2$ test statistic is used to perform the fit between two MC templates, defined as:
\begin{equation}\label{eq:mod_chi2}
	\chi^2_{\mathrm{mod}} = \sum_{i \in \mathrm{bins}}^{}\frac{(N^{\mathrm{exp}}_i - N^{\mathrm{obs}}_i)^2}{N^{\mathrm{exp}}_i + (\sigma^{\mathrm{exp}}_i)^2} + \sum_{j \in \mathrm{syst}}^{}\frac{(s_j - \hat{s_j})^2}{\sigma^2_{s_j}} \,,
\end{equation}
where $N^{\mathrm{exp}}_i$ and $N^{\mathrm{obs}}_i$ are the total number of expected and observed events in the $i^{th}$ bin, respectively. The denominator in the first term of Eq.~\ref{eq:mod_chi2} accounts for both the Poisson error in the expected events $\sqrt{N^{\mathrm{exp}}_i}$ and the additional uncertainty due to the limited MC statistics $\sigma^{\mathrm{exp}}_i$ in the $i^{th}$ bin. The second term of Eq.~\ref{eq:mod_chi2} represents the penalty for deviation of systematic parameters from their prior expectations. Here, $\hat{s_j}$ is the best-fit value of the $j^{th}$ nuisance parameter, $s_j$ is its nominal value, and $\sigma_{s_j}$ is the standard deviation of its Gaussian prior.

\subsection{Systematic uncertainties}
\label{sec:sys_uncertanity}

Nuisance parameters are included in the analysis to account for uncertainties that can affect both the shape and amplitude of the expected event distributions. These systematic uncertainties are categorized based on their physical origin, as summarized in Table~\ref{tab:systematic_params}.

\begin{table}[ht]
	\centering
	\begin{tabular}{lcc}
		\hline
		\hline
		\textbf{Parameter} & \textbf{Nominal value} & \textbf{Prior width} \\
		\hline
		\multicolumn{3}{@{}l}{\textbf{Detector:}} \\
		DOM efficiency IC86       & 1.0 & $\pm\,$0.1 \\
		DOM efficiency ICU        & 1.0 & $\pm\,$0.05 \\
		Ice absorption            & 1.0 & [0.9, 1.1] \\
		Ice scattering            & 1.05 & [0.95, 1.15] \\
		Relative eff. $p_0$       & 0.10 & [$-\,0.6$, 0.5] \\
		Relative eff. $p_1$       & $-\,0.05$ & [$-\,0.2$, 0.2] \\
		\hline
		\multicolumn{3}{@{}l}{\textbf{Atmospheric flux:}} \\
		$\Delta \gamma_{\nu}$     & 0.0 & $\pm\,$0.1 \\
		$\Delta \pi^+ \text{ yields D}$ & 0.0 & $\pm\,$0.3 \\
		$\Delta \pi^+ \text{ yields G}$ & 0.0 & $\pm\,$0.3 \\
		$\Delta \pi^+ \text{ yields H}$ & 0.0 & $\pm\,$0.15 \\
		$\Delta \pi^+ \text{ yields I}$ & 0.0 & $\pm\,$0.122 \\
		$\Delta K^+ \text{ yields W}$ & 0.0 & $\pm\,$0.4 \\
		$\Delta K^+ \text{ yields Y}$ & 0.0 & $\pm\,$0.3 \\
		$\Delta K^+ \text{ yields Z}$ & 0.0 & $\pm\,$0.122 \\
		\hline
		\multicolumn{3}{@{}l}{\textbf{Cross-section:}} \\
		$M_A^{\text{CCQE}}$ (in $\sigma$)   & 0.0 & $\pm\,$1.0 \\
		$M_A^{\text{CCRES}}$ (in $\sigma$) & 0.0 & $\pm\,$1.0 \\
		$M_A^{\text{NCRES}}$ (in $\sigma$) & 0.0 & $\pm\,$1.0 \\
		$M_A^{\text{coh}}$ (in $\sigma$)   & 0.0 & $\pm\,$1.0 \\
		DIS CSMS                 & 0.0 & $\pm\,$1.0 \\
		$\nu_\tau$ x-sec scale   & 0.0 & [$-\,1.0$, $+\,1.0$] \\
		\hline
		\multicolumn{3}{@{}l}{\textbf{Normalization:}} \\
		$A_{\text{eff}}$ scale    & 1.0 & [0.1, 2.0] \\
		\hline
		\multicolumn{3}{@{}l}{\textbf{Atmospheric muons:}} \\
		Atm. $\mu$ scale          & 1.0 & [0.1, 3.0] \\
		\hline
		\multicolumn{3}{@{}l}{\textbf{Oscillations:}} \\
		$\theta_{23}$            & 47.50$^\circ$ & [38$^\circ$, 52$^\circ$] \\
		$\Delta m^2_{31}$        & 0.00247 eV$^2$ & [0.001, 0.004] eV$^2$ \\
		\hline
		\hline
	\end{tabular}
	\caption{Summary of systematic uncertainty parameters in terms of their nominal values. For parameters with Gaussian priors, the listed values correspond to the $\pm \, 1\sigma$ prior range, while for parameters with uniform priors, the full allowed range is listed.}
	\label{tab:systematic_params}
\end{table}

Atmospheric neutrino flux uncertainties include the spectral index of the primary cosmic rays and hadronic interaction uncertainties, following the work from Ref.~\cite{Barr:2006it}. The neutrino flux prediction from Ref.~\cite{Honda:2015fha} is used as the baseline model. Neutrino-nucleon cross section uncertainties are treated using the reweighting framework provided by GENIE. For quasielastic (QE) and resonance (RES) interactions, uncertainties are parameterized by varying the axial mass parameter. At higher energies, where deep inelastic scattering (DIS) dominates, an interpolated model between GENIE and the CSMS~\cite{Cooper-Sarkar:2011jtt} prediction is used. Detector-related uncertainties are addressed using the likelihood-free inference method introduced in~\cite{Fischer:2023dbo}, which parametrizes variations in the optical properties of the Antarctic ice, such as absorption and scattering lengths, as well as the efficiency of the optical modules used in both IceCube DeepCore and the IceCube Upgrade. Additionally, two separate parameters ($p_0$ and $p_1$)~\cite{IceCube:2023ahv} are included to account for the angular acceptance of the optical module. Finally, independent normalization parameters are assigned to atmospheric muons and neutrinos. As far as neutrino oscillation parameters are concerned, the impact of the solar ($\theta_{12}$ and $\Delta m^2_{21}$) and reactor ($\theta_{13}$) parameters, within their present uncertainties, is found to be negligible for these analyses. Therefore, their values are fixed to those from the latest global fit (NuFit 5.2~\cite{Esteban:2020cvm}), as listed in Table~\ref{tab:osc-param-value}. Furthermore, the $\delta_{\text{CP}}$-dependent terms in $P(\nu_e \rightarrow \nu_\mu)$, $P(\nu_e \rightarrow \nu_\tau)$, $P(\nu_\mu \rightarrow \nu_\mu)$, and $P(\nu_\mu \rightarrow \nu_\tau)$ are suppressed by a factor of $(\Delta m^2_{21} / \Delta m^2_{31})\sin\theta_{13}$, which is about 0.005 (see Eqs. 3.5, 3.6, 3.7, and 3.8) in Ref.~\cite{Akhmedov:2004ny}. Therefore, the effect of $\delta_{\text{CP}}$ on atmospheric neutrino oscillations is negligible for GeV-energy neutrinos traversing long baselines. We have checked the effect of the minimization and true choices of $\delta_\text{CP}$ on sensitivities of our studies and found them to be within a few percent; hence, we fix $\delta_\text{CP}=0$. Only the atmospheric oscillation parameters ($\theta_{23}$ and $\Delta m^2_{31}$) are treated as free parameters in the fit. All of the nuisance parameters discussed above are described in detail in Refs.~\cite{IceCubeCollaboration:2023wtb,IceCubeCollaboration:2024ssx,IceCube:2025chb}.

\subsection{Statistical approach}
\label{sec:statistical_approch}

Rejecting the vacuum or uniform density profile with respect to the 12-layered PREM profile involves a binary hypothesis test comparing two non-nested hypotheses. This differs from the case of nested hypothesis testing, where a general model is tested against a specific one that can be obtained by a certain realization of the parameters of general model. In nested hypothesis tests, {\it Wilks' theorem}~\cite{Wilks:1938dza} can be applied to derive sensitivities. In contrast, the Wilks' theorem does not apply to non-nested hypothesis testing, as the models under comparison are discrete and not connected through a continuous variation of parameters.

The sensitivity to reject the test hypothesis in favor of the true hypothesis, expressed in units of Gaussian sigma, is calculated following Ref.~\cite{Ciuffoli:2013rza}, and is given by:
\begin{equation}
	\eta_\sigma = \frac{\Delta \chi^2_\text{mod} (\text{test}) - \Delta \chi^2_\text{mod} (\text{true})}{2\sqrt{\Delta \chi^2_\text{mod} (\text{test})}} \,,
	\label{eq:sen_non_nested}
\end{equation}
where $\Delta \chi^2_\text{mod} (\text{true})$ is defined as the difference between the metric values,
\begin{equation}
	\Delta \chi^2_\text{mod} (\text{true}) = \chi^2_\text{mod} (\text{true}|\text{true}) - \chi^2_\text{mod} (\text{test}|\text{true})\,.
\end{equation}

For $\Delta \chi^2_\text{mod} (\text{true})$, the Asimov dataset is generated assuming the true hypothesis, which is fitted with the test hypothesis to obtain $\chi^2_\text{mod}(\text{test}|\text{true})$ and with the true hypothesis to obtain $\chi^2_\text{mod} (\text{true}|\text{true})$. By construction, $\chi^2_\text{mod} (\text{true}|\text{true}) = 0$, since the Asimov dataset is generated without statistical fluctuations. The fit performed with the test hypothesis, while obtaining $\chi^2_\text{mod} (\text{test}|\text{true})$, yields a set of best-fit parameters, which are then used to generate another Asimov dataset under the test hypothesis. This second Asimov dataset is subsequently fitted using the true hypothesis to obtain $\chi^2_\text{mod} (\text{true}|\text{test})$, and using the test hypothesis to obtain $\chi^2_\text{mod} (\text{test}|\text{test})$. The difference of these two metric values is used to define the second test-statistic, $\Delta \chi^2_\text{mod} (\text{test})$:
\begin{equation}
	\Delta \chi^2_\text{mod} (\text{test}) = \chi^2_\text{mod} (\text{true}|\text{test}) - \chi^2_\text{mod} (\text{test}|\text{test})\,.
\end{equation}

On the other hand, the mass of the Earth can be estimated by uniformly scaling the densities of all layers in the 12-layered PREM profile using a continuous parameter, denoted by $\alpha$. Similarly, for correlated density measurement, variations in the densities of different layers of the 5-layered PREM profile are parameterized using a single continuous scaling parameter, denoted by $\alpha_\text{C}$. Both these scenarios correspond to nested hypothesis testing, where the Wilk's theorem is used to evaluate confidence levels from $\Delta \chi^2_\text{mod}$, which is given as, 
\begin{equation}
	\Delta \chi^2_\text{mod} = \chi^2_\text{mod} (\alpha') - \chi^2_\text{mod} (\alpha'_0)\,,
\end{equation}
where $\alpha'$ represents the parameter $\alpha$ for the mass of the Earth and the parameter $\alpha_\text{C}$ for the correlated density measurement. The Asimov data is generated with a given value of the scaling parameter $\alpha' = \alpha'_0$, and fitted using the same model by scanning over $\alpha'$.	

This analysis considers two data-taking scenarios for sensitivity projections. The first assumes that the current IceCube DeepCore configuration, IC86, continues data collection for a total of 15 years. This scenario is referred to as ``IC86 (15 yr)''. The second scenario involves a joint fit combining 12 years of IC86 data with an additional 3 years from the IceCube Upgrade configuration, IC93, beginning in 2026. This scenario is denoted as ``IC86 (12 yr) + IC93 (3 yr)''. Note that the scenario corresponding to the currently operating IceCube Upgrade lies between these two extreme cases.

\section{Sensitivity results}
\label{sec:sensitivity_result}

This section presents the expected sensitivities of the study using the IceCube Upgrade detector. The sensitivities for rejecting the vacuum and uniform density hypotheses are shown as a function of the true value of $\sin^2\theta_{23}$, since $\theta_{23}$ remains one of the least precisely known oscillation parameters. In the fit, both $\theta_{23}$ and $\Delta m^2_{31}$ are treated as free parameters, along with several nuisance parameters (as detailed in section~\ref{sec:sys_uncertanity}). The remaining oscillation parameters, $\theta_{12}$, $\Delta m^2_{21}$, and $\theta_{13}$, are fixed to the values given in Table~\ref{tab:osc-param-value}, due to their precise experimental measurements and minor impact on the analysis. We also fix $\delta_{\text{CP}}$ to zero, as its impact on these analyses is minimal. All sensitivity result plots shown in this section assume NO. Sensitivities under IO assumption have also been evaluated and found to be lower than those for NO. This happens because the matter effect is mainly experienced by antineutrinos, that have lower statistics due to their smaller flux and cross sections. Note that the sensitivities discussed in this section are obtained assuming the originally planned IC93 configuration. Since the detector is currently operating in the IC91 configuration, the actual sensitivity is expected to be lower than the values reported here.

\subsection{Sensitivity to Establishing Earth Matter Effects}
\label{sec:analysis_i_result}

\begin{figure}[htp!]
	\includegraphics[width=\linewidth]{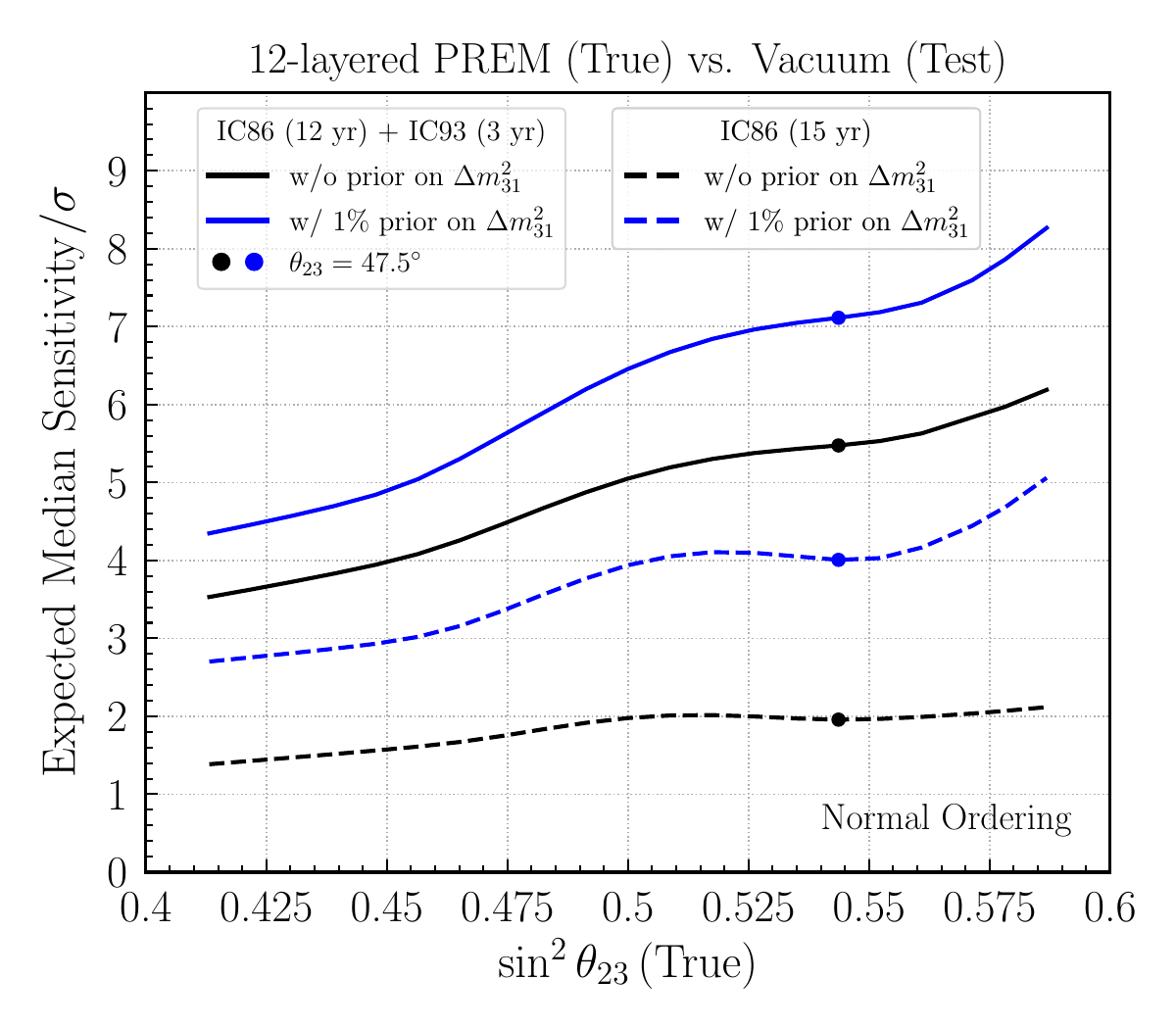}
	\caption{Expected median sensitivity as a function of the true value of $\sin^2\theta_{23}$ for establishing Earth matter effects in atmospheric neutrino oscillations, by rejecting the vacuum oscillation hypothesis with respect to the 12-layered PREM hypothesis. The blue (black) curves show the sensitivity with (without) a 1\% prior on $\Delta m_{31}^2$. The dashed curves correspond to the sensitivity from the current DeepCore configuration with 15 years of exposure, while the solid curves represent the sensitivity from a joint fit using 12 years of DeepCore and 3 years of Upgrade simulated data.}
	\label{fig:result_analysis_i_prior}
\end{figure}

The expected sensitivity to establish Earth matter effects is presented by quantifying the significance with which the vacuum oscillation hypothesis can be rejected with respect to the matter oscillation hypothesis. Figure~\ref{fig:result_analysis_i_prior} shows the expected median sensitivity to reject the vacuum oscillation hypothesis as a function of the true value of $\sin^2\theta_{23}$, assuming NO. The solid black curve represents the sensitivity when 3 years of IceCube Upgrade (IC93) simulated data are combined with 12 years of IceCube DeepCore (IC86) simulated data. The dashed black curve shows the sensitivity assuming that the current IceCube DeepCore setup continues data taking for 15 years. This comparison highlights the sensitivity enhancement provided by the IceCube Upgrade, which improves the significance by approximately a factor of two to three depending on the true value of $\theta_{23}$ and a prior on $\Delta m^2_{31}$ compared to IceCube DeepCore alone. The circular markers on each curve indicate the sensitivities for a representative value of $\theta_{23} = 47.5^\circ$ (best-fit value reported by the IceCube DeepCore measurement~\cite{IceCubeCollaboration:2024ssx}), yielding a significance of $5.5\sigma$ for the combined fit using IC93 and IC86 and $1.9\sigma$ for IC86 alone in the case of NO. For IO, the significance for the combined fit reduces to 3.4$\sigma$.

So far, the study has examined how the sensitivity of this analysis depends on the true value of $\theta_{23}$. However, it is also observed that allowing $\Delta m^2_{31}$ to vary freely in the fit reduces the sensitivity to matter effects measurement. This happens due to the degeneracy between $\Delta m^2_{31}$ and matter effects, where variations in $\Delta m^2_{31}$ can shift the location of the oscillation valley in a way that mimics the variation introduced by matter effect. To reduce the impact of this degeneracy, external constraint on $\Delta m^2_{31}$ is incorporated using a Gaussian prior. This approach is motivated by the fact that $\Delta m^2_{31}$ is already measured with a 1$\sigma$ precision of approximately 1\% as reported by NuFit~5.2~\cite{Esteban:2020cvm} (excluding Super-Kamiokande data), and Jiangmen Underground Neutrino Observatory (JUNO)~\cite{JUNO:2015zny} is expected to improve this precision to about 0.5\%~\cite{JUNO:2022mxj}. In particular, a Gaussian prior on $\Delta m^2_{31}$ is included through the pull-penalty term of Eq.~\ref{eq:mod_chi2} for the minimization of $\chi^2_\text{mod}$. The prior is centered at the nominal value of $2.47 \times 10^{-3}~\text{eV}^2$ with a width ($\sigma_{\Delta m^2_{31}}$) equivalent to $2.47 \times 10^{-5}~\text{eV}^2$, corresponding to 1\% of its nominal value. To illustrate the role of this prior in reducing the above mentioned degeneracy, Fig.~\ref{fig:result_analysis_i_prior} presents the expected sensitivity to reject the vacuum oscillation hypothesis as a function of the true value of $\sin^2\theta_{23}$ with (blue curves) and without (black curves) the prior on $\Delta m^2_{31}$. Solid curves correspond to the joint fit of IC93 and IC86, while dashed curves correspond to IC86 alone. All sensitivity curves are obtained assuming normal mass ordering. It is observed that including the prior on $\Delta m^2_{31}$ enhances the expected sensitivity by approximately 30\%. This improvement occurs because incorporating external information on $\Delta m^2_{31}$ reduces its degeneracy with matter effects, allowing the matter-induced features to be identified more efficiently. As a result, the significance of establishing the presence of matter effects is greatly enhanced. The circular markers on blue curves indicate the sensitivities for a representative value of $\theta_{23} = 47.5^\circ$, yielding a significance of $7.1\sigma$ for the combined fit using IC93 and IC86, and $4.0\sigma$ for IC86 alone.

The approximately linear behavior of the sensitivity curves in Fig.~\ref{fig:result_analysis_i_prior} as a function of $\sin^2\theta_{23}$ arises from the fact that the dominant contributions of matter effects to the $\nu_\mu \rightarrow \nu_\mu$ survival probability and the $\nu_e \rightarrow \nu_\mu$ appearance probability are proportional to $\sin^2\theta_{23}$, as demonstrated through a series expansion in Ref.~\cite{Akhmedov:2004ny}.

\subsection{Sensitivity to Validate the Layered Structure}
\label{sec:analysis_ii_result}

\begin{figure}[htp!]
	\includegraphics[width=\linewidth]{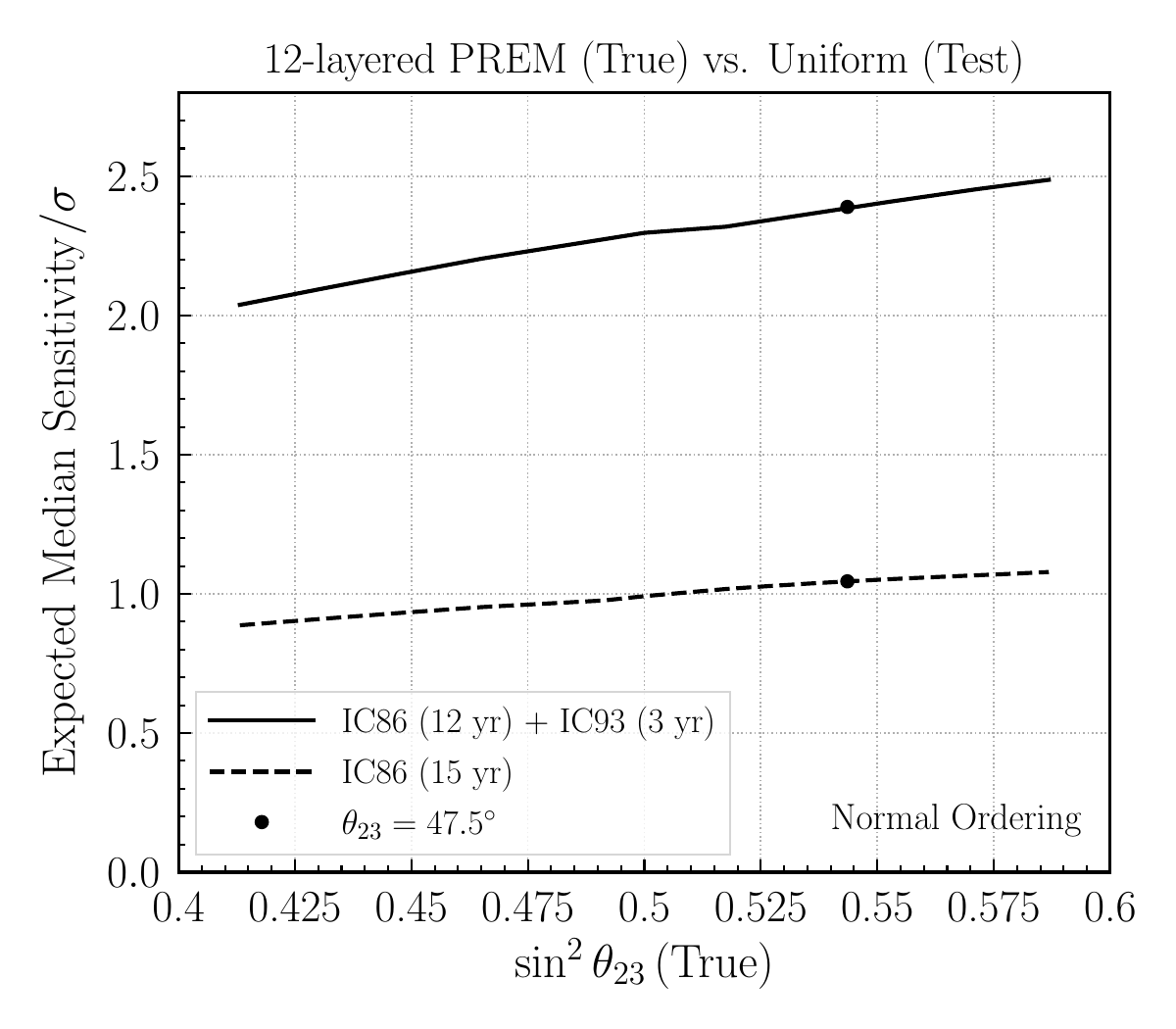}
	\caption{Expected median sensitivity as a function of the true value of $\sin^2\theta_{23}$ for validating the layered structure of the Earth by rejecting the uniform matter density profile with respect to the 12-layered PREM density profile. The dashed curve shows the sensitivity using the current DeepCore configuration with 15 years of exposure, while the solid curve corresponds to the sensitivity from a joint fit using 12 years of DeepCore and 3 years of Upgrade simulated data.}
	\label{fig:result_analysis_ii} 
\end{figure}

This section presents the expected sensitivity to validate the layered structure of the Earth by estimating the significance with which the uniform density hypothesis can be rejected with respect to the 12-layered PREM density profile. Figure~\ref{fig:result_analysis_ii} shows the expected median sensitivity to reject the uniform hypothesis as a function of the true value of $\sin^2\theta_{23}$, assuming NO. The solid curve represents the sensitivity from the combined fit using IC93 and IC86, while the dashed curve corresponds to IC86 alone. For a representative choice of $\theta_{23} = 47.5^\circ$, the sensitivity reaches approximately $2.4\sigma$ for the combined IC93 and IC86 fit, and about $1.0\sigma$ for IC86 alone, under the assumption of NO. For IO, the sensitivity to reject uniform hypothesis for the combined fit is 1.4$\sigma$.

Unlike the previous analysis, considering a prior on $\Delta m^2_{31}$ has minimal impact on the sensitivity in this case; therefore, we do not present separate sensitivity results with prior. The approximately linear dependence of the sensitivity curves on $\sin^2\theta_{23}$ arises for the same reasons discussed in Section~\ref{sec:analysis_i_result}.

\subsection{Sensitivity to Measure the Mass of the Earth}
\label{sec:result_analysis_iii_A}

\begin{figure}[htp!]
	\includegraphics[width=\linewidth]{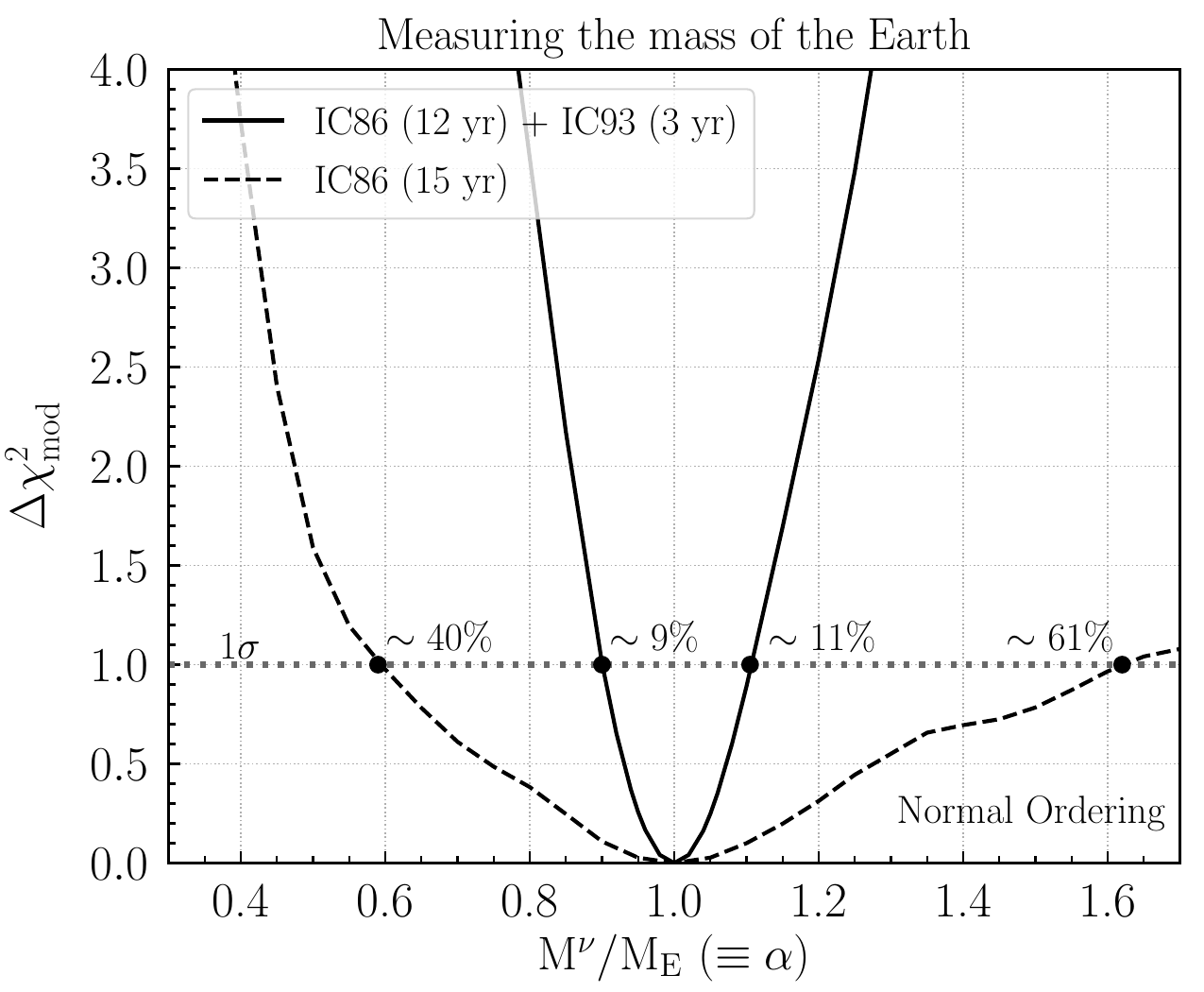}
	\caption{Expected $\Delta \chi^2_\text{mod}$  as a function of the scaling parameter $\alpha$ for measuring the mass of the Earth using atmospheric neutrino oscillations. The dashed curve shows the sensitivity obtained with the current DeepCore configuration assuming 15 years of exposure, while the solid curve represents the sensitivity from a joint fit using 12 years of DeepCore and 3 years of Upgrade simulated data.}
	\label{fig:result_analysis_iii_mass}
\end{figure}

This section presents the expected median sensitivity to measure the mass of the Earth using oscillating neutrinos. Figure~\ref{fig:result_analysis_iii_mass} shows $\Delta \chi^2_\text{mod}$ as a function of the scaling parameter, $\alpha$, assuming normal mass ordering. Here, $\alpha = 1.0$ corresponds to the mass of the Earth as determined from gravitational measurements, while values of $\alpha$ less than or greater than 1.0 correspond to lighter or heavier Earth models, respectively. The solid curve represents the sensitivity from a joint fit using both IC86 and IC93 simulated MC data, while the dashed curve shows the sensitivity using IC86 alone. The results indicate that with 15 years of IC86 exposure, the mass of the Earth can be constrained to approximately 50\% at $1\sigma$ precision. However, in the scenario of combined fit, the precision\footnote{The precision is defined as $(\alpha'_\text{max}-\alpha'_\text{min})/2$, where $\alpha'_\text{max}$ and $\alpha'_\text{min}$ are the maximum and minimum values of $\alpha'$ at a given confidence level.} improves to about 10\% at the $1\sigma$ level. In the case of combined fit under the assumption of IO, the $1\sigma$ precision drops to $\sim$17\%.

\subsection{Sensitivity to Correlated Density Measurement}
\label{sec:result_analysis_iii_B}

\begin{figure*}[htp!]
	\centering
	
	\begin{minipage}{0.47\textwidth}
		\centering
		\includegraphics[width=\linewidth]{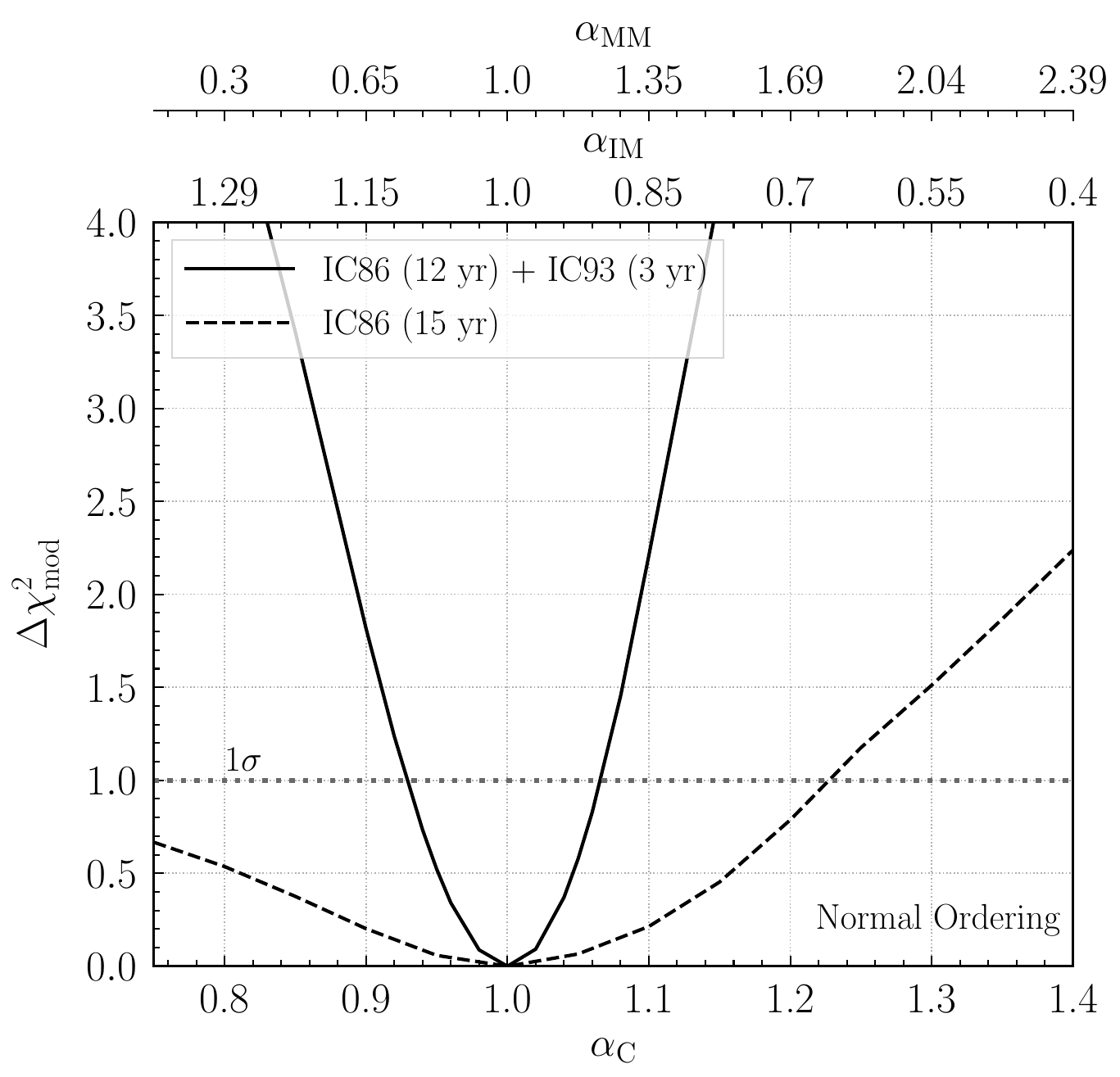}
		
		\vspace{0.5em}
		\centerline{(a)}
	\end{minipage}
	\begin{minipage}{0.5\textwidth}
		\centering
		\vspace{3.5em}
		\includegraphics[width=\linewidth]{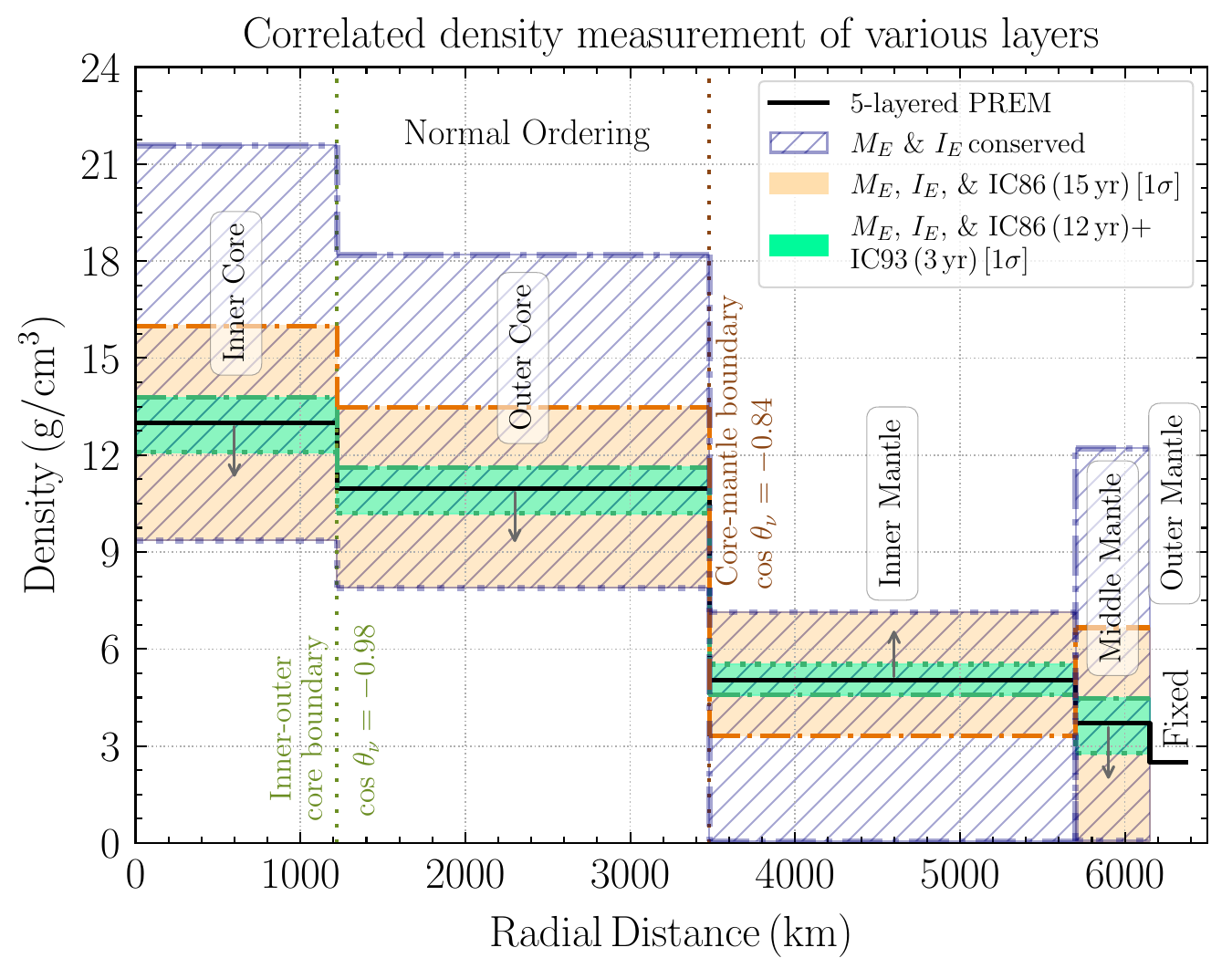}
		\vspace{0.5em}
		\centerline{(b)}
	\end{minipage}
	
	\caption{(a) Expected $\Delta \chi^2_\text{mod}$ as a function of the individual scaling 	factors for correlated density measurement using atmospheric neutrino oscillations. The three x-axes correspond to the scaling factors for the core ($\rm{\alpha_C}$), inner mantle ($\rm{\alpha_{IM}}$), and middle mantle ($\rm{\alpha_{MM}}$). The dashed curve shows the sensitivity obtained with 15 years of exposure using the current IceCube DeepCore configuration. The solid line shows the sensitivity obtained from the joint fit using 12 years of IceCube DeepCore and 3 years of IceCube Upgrade. (b) The $1\sigma$ allowed density bands derived from atmospheric neutrino oscillation data combined with external geophysical constraints. The hatched region represents the range of densities permitted solely by the constraints on the total mass ($M_E$) and moment of inertia ($I_E$) of the Earth. The orange band shows the $1\sigma$ bounds from a fit using 15 years of simulated DeepCore data with these external constraints. The green band corresponds to the $1\sigma$ bounds from a joint fit using 12 years of DeepCore data and 3 years of Upgrade data, combined with the same geophysical constraints. The boundaries of the $1\sigma$ regions are indicated by the orange and green lines, respectively. }
	\label{fig:result_analysis_iii_core}
\end{figure*}

This section discusses the expected median sensitivity for the measurement of the correlated densities of the different layers inside the Earth using a 5-layered PREM profile. In this study, external constraints, such as the total mass and moment of inertia of the Earth, lead to correlations among the densities of different layers. This occurs because the variation in each layer’s density is not independent but must collectively satisfy these global constraints. This allows the sensitivity curve to be interpreted in terms of any of the three scaling factors - $\rm{\alpha_C}$ (core), $\rm{\alpha_{IM}}$ (inner mantle), and $\rm{\alpha_{MM}}$ (middle mantle). Figure~\ref{fig:result_analysis_iii_core}(a) shows $\Delta \chi^2_\text{mod}$ as a function of these three scaling factors - $\rm{\alpha_C}$, $\rm{\alpha_{IM}}$,  and $\rm{\alpha_{MM}}$. Each of the three x-axes corresponds to a different scaling factor, and any one of them can be chosen as the observable, while the others can be interpreted in terms of it. In this analysis, $\rm{\alpha_C}$ has been treated as the physics parameter.  The solid black curve shows the sensitivity when 12 years of IC86 and 3 years of IC93 simulated neutrino data are combined with the external constraints. The dashed black curve shows the sensitivity when 15 years of simulated IC86 neutrino data is combined with the external constraints. The precisions in the estimation of the correlated densities of the different layers improve remarkably for the combined fit, reaching about $\sim\,$7\% for the core, or $\sim\,$10\% for the inner mantle, or $\sim\,$23\% for the middle mantle. On the other hand,  for IO, we obtain a precision of $\sim\,$12\% for the core, or $\sim\,$18\% for the inner mantle, or $\sim\,$42\% for middle mantle from the combined fit.

Figure~\ref{fig:result_analysis_iii_core}(b) presents the allowed density ranges for different layers of the Earth, based on the 5-layered PREM profile. This figure is analogous to  Fig.~\ref{fig:radial_density_profile}, where the black curve represents the density values of the 5-layered PREM profile as a function of radial distance from the center of the Earth. The white-hatched region indicates the density ranges allowed by the external constraints of the total mass ($M_E$) and moment of inertia ($I_E$) of the Earth. The orange band shows the $1\sigma$ allowed density ranges when these gravitational constraints are combined with 15 years of simulated IC86 data. This band in Fig.~\ref{fig:result_analysis_iii_core}(b) corresponds to the $1\sigma$ bounds obtained from the dashed black curve in Fig.~\ref{fig:result_analysis_iii_core}(a). A one-sided bound is observed in Fig.~\ref{fig:result_analysis_iii_core}(a), which represents the upper limits for the inner core, outer core, and middle mantle, and at the same time, it corresponds to a lower limit for the inner mantle. In other words, the densities of the inner and outer core are positively correlated with the middle mantle, but negatively correlated with the inner mantle. This behavior highlights the intricate correlations among the densities of the different layers.  The arrows in Fig.~\ref{fig:result_analysis_iii_core}(b) highlight these correlations between different Earth layers. The green band shows the improved $1\sigma$ constraints obtained from a joint fit using 12 years of simulated IC86 data and 3 years of simulated IC93 data. This green band has been obtained from the $1\sigma$ bounds of the solid black curve in Fig.~\ref{fig:result_analysis_iii_core}(a). The plot illustrates a significant reduction in the allowed parameter space when IceCube Upgrade data are included alongside the external constraints. No uncertainty bands are shown for the outermost layer, as its density is kept fixed in the model due to its well-established value.

These sensitivity results demonstrate that atmospheric neutrino oscillation data can further constrain the allowed density of different layers of the Earth beyond what is possible with mass and moment of inertia measurements alone, highlighting the complementarity of neutrino oscillation based Earth tomography with traditional geophysical methods.

\subsection{Impact of Systematic Uncertainties on Sensitivities}
\label{sec:impact_sys_uncer}

\begin{figure}[htp!]
	\includegraphics[width=\linewidth]{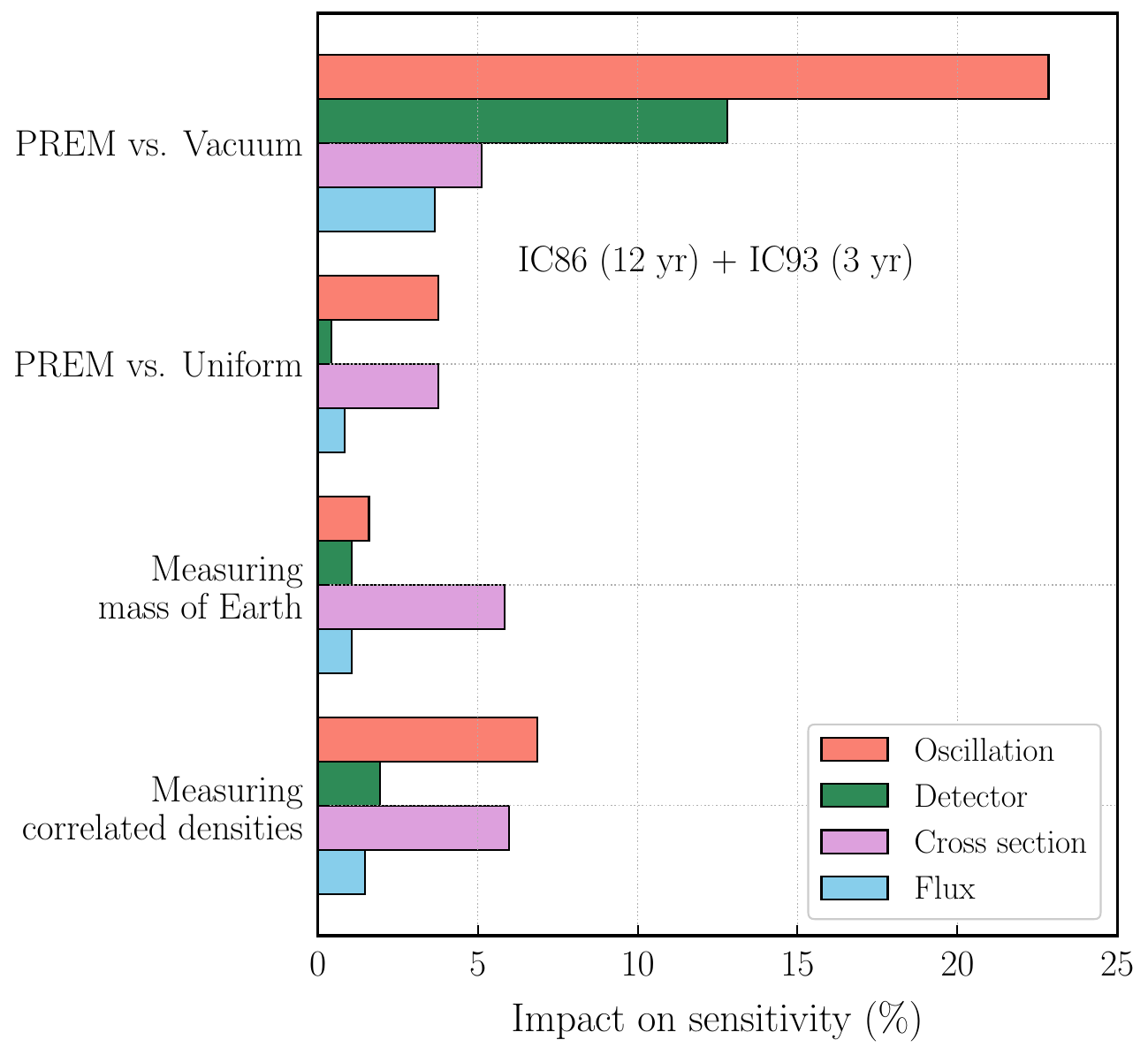}
	\caption{Relative impact of different groups of systematic uncertainties on the sensitivity, in terms of the number of $\sigma$, for each analysis considered in this study, assuming perfect knowledge of one group at a time. For the analyses PREM vs. Vacuum and PREM vs. Uniform, the sensitivities  are calculated for a benchmark value of $\theta_{23}=47.5^{\circ}$. For the analysis of the mass of the Earth (correlated density measurement), the sensitivities correspond to the scaling parameter  $\alpha=0.8$ ($\alpha_{\rm C}=0.8$) in fit and $\alpha=1.0$ ($\alpha_{\rm C}=1.0$) in truth. The horizontal bars represent the change in sensitivity in percentage when the corresponding systematic group is fixed as compared to when it is free. The y-axis lists the four analyses performed in this work.}
	\label{fig:impact_sys_uncer}
\end{figure}

To understand how different groups of systematic uncertainty parameters affect the sensitivity projections of each analysis, the relative impact on sensitivity is evaluated by assuming that all nuisance parameters within a given systematic group are perfectly known. Figure~\ref{fig:impact_sys_uncer} illustrates the percentage change in sensitivity when the nuisance parameters within each group are fixed during the minimization process for the joint fit of IC86 (12 yr) + IC93 (3 yr). The relative impact on sensitivity is quantified as:
\begin{equation}
	\text{Impact on sensitivity (\%)} = \frac{\sigma^\text{fixed} - \sigma^\text{free}}{\sigma^\text{free}} \times 100 \,,
	\label{eq:impact_sys_uncer}
\end{equation}
where $\sigma^\text{fixed}$ denotes the sensitivity obtained when all nuisance parameters within a specific systematic group are fixed, while $\sigma^\text{free}$ in Eq.~\ref{eq:impact_sys_uncer} represents the sensitivity when all systematic parameters listed in Table~\ref{tab:systematic_params} are allowed to vary freely during minimization. Additionally, $\sigma^\text{free}$ for the ``PREM vs. Vacuum'' analysis is considered w/o a prior on $\Delta m^2_{31}$.

Among all the systematic groups considered, the oscillation parameters have the most significant impact on the sensitivity to establish matter effects (PREM vs.~Vacuum), whereas detector, flux, and cross section systematics contribute relatively less. This is already evident in Fig.~\ref{fig:result_analysis_i_prior}, where applying an external Gaussian prior on $\Delta m^2_{31}$ leads to a noticeable enhancement in sensitivity. In contrast, for the validating layered structure analysis (PREM vs. Uniform), the impacts of oscillation and cross section systematics are comparable and both remain below 5\%, with detector and flux systematics having a minor effect. For the case of measuring the total mass of the Earth, cross section uncertainties exhibit a relatively larger effect, comparable to the contributions from other systematic groups, but still remain below the 10\% level. Finally, for correlated density measurement, flux and cross section systematics have a moderate and comparable impact, each contributing less than 10\%, while detector and oscillation systematics remain subdominant.

Figure~\ref{fig:impact_sys_uncer} highlights the potential for substantial improvement in sensitivity through more precise measurements of the oscillation and other systematic parameters.

\section{Summary and conclusion}
\label{sec:conclusion}

In this study, we investigate the potential of the IceCube Upgrade detector to establish matter effects in atmospheric neutrino oscillations by rejecting the vacuum oscillation hypothesis with respect to the matter oscillation hypothesis. Utilizing these matter effects, we further project the sensitivity of the IceCube Upgrade to validate the layered structure inside the Earth, measure the mass of the Earth, and constrain the correlated densities of different layers within the Earth.

The addition of new strings equipped with advanced optical modules will significantly increase the detection rate of GeV-scale neutrinos, improve event reconstruction, and improve particle identification. The deployment of various calibration devices on these strings will further help reduce systematic uncertainties. These improvements lead to a substantial enhancement in sensitivity for matter-effect-driven neutrino oscillation analyses. As demonstrated in this study, all our analyses show improved sensitivities compared to scenarios without these additional strings. Moreover, it is observed that for the joint fit combining 3 years of IC93 simulated data with 12 years of IC86 simulated data, the IceCube Upgrade (IC93) contributes dominantly to the sensitivity across all analyses compared to IceCube DeepCore (IC86).

With the advancement in the precision measurement of neutrino oscillation parameters, neutrino oscillation tomography has emerged as a promising probe of the interior of the Earth. Information obtained through weak interactions of neutrinos provides a complementary and independent probe compared to traditional methods such as seismic studies and gravitational measurements. The present study aims to establish a robust methodology for future Earth tomography analyses using the IceCube Upgrade detector.

\vskip0.5cm

\noindent {\it Note added.} Sensitivities in this work assume the originally planned seven-string configuration for the Upgrade (IC93). The 2025-2026 deployment resulted in a reduced configuration. While full performance studies for the as-built detector are ongoing, the Upgrade is expected to retain a significant sensitivity advantage over the IC86-only case discussed in this manuscript. Nevertheless, the smaller number of strings and optical modules in the currently operating IC91 configuration will lead to reduced sensitivity relative to the IC93 baseline. These estimates are conservative, as no reoptimization was performed for the reduced geometry, and the approximately uniform nature of the degradation implies that additional livetime would be needed to reach the sensitivities presented in this paper.

\begin{acknowledgements}
	The IceCube Collaboration acknowledges the significant contributions to this manuscript from Sharmistha Chattopadhyay, Krishnamoorthi J, and Anuj Kumar Upadhyay. The authors gratefully acknowledge the support from the following agencies and institutions:
	USA {\textendash} U.S. National Science Foundation-Office of Polar Programs,
	U.S. National Science Foundation-Physics Division,
	U.S. National Science Foundation-EPSCoR,
	U.S. National Science Foundation-Office of Advanced Cyberinfrastructure,
	Wisconsin Alumni Research Foundation,
	Center for High Throughput Computing (CHTC) at the University of Wisconsin{\textendash}Madison,
	Open Science Grid (OSG),
	Partnership to Advance Throughput Computing (PATh),
	Advanced Cyberinfrastructure Coordination Ecosystem: Services {\&} Support (ACCESS),
	Frontera and Ranch computing project at the Texas Advanced Computing Center,
	U.S. Department of Energy-National Energy Research Scientific Computing Center,
	Particle astrophysics research computing center at the University of Maryland,
	Institute for Cyber-Enabled Research at Michigan State University,
	Astroparticle physics computational facility at Marquette University,
	NVIDIA Corporation,
	and Google Cloud Platform;
	Belgium {\textendash} Funds for Scientific Research (FRS-FNRS and FWO),
	FWO Odysseus and Big Science programmes,
	and Belgian Federal Science Policy Office (Belspo);
	Germany {\textendash} Bundesministerium f{\"u}r Bildung und Forschung (BMBF),
	Deutsche Forschungsgemeinschaft (DFG),
	Helmholtz Alliance for Astroparticle Physics (HAP),
	Initiative and Networking Fund of the Helmholtz Association,
	Deutsches Elektronen Synchrotron (DESY),
	and High Performance Computing cluster of the RWTH Aachen;
	Sweden {\textendash} Swedish Research Council,
	Swedish Polar Research Secretariat,
	Swedish National Infrastructure for Computing (SNIC),
	and Knut and Alice Wallenberg Foundation;
	European Union {\textendash} EGI Advanced Computing for research;
	Australia {\textendash} Australian Research Council;
	Canada {\textendash} Natural Sciences and Engineering Research Council of Canada,
	Calcul Qu{\'e}bec, Compute Ontario, Canada Foundation for Innovation, WestGrid, and Digital Research Alliance of Canada;
	Denmark {\textendash} Villum Fonden, Carlsberg Foundation, and European Commission;
	New Zealand {\textendash} Marsden Fund;
	Japan {\textendash} Japan Society for Promotion of Science (JSPS)
	and Institute for Global Prominent Research (IGPR) of Chiba University;
	Korea {\textendash} National Research Foundation of Korea (NRF);
	Switzerland {\textendash} Swiss National Science Foundation (SNSF); India {\textendash} Department of Atomic Energy (DAE), Department of Science and Technology (DST), and  Anusandhan National Research Foundation (ANRF).
\end{acknowledgements}

\appendix
\vspace{1cm}

\section{Differences in various neutrino oscillograms for different Earth density profiles}
\label{app:prob}

This appendix presents the impact of different Earth density profiles on neutrino oscillation probabilities in ($E_\nu$, $\cos\theta_\nu$) plane. To illustrate these effects, the probability differences between alternative Earth density profiles and the 12-layered PREM density profile are presented. The probabilities for each density profile are calculated using the nominal values of oscillation parameters given in Table~\ref{tab:osc-param-value}. Figure~\ref{fig:prem_osc} of Section~\ref{sec:matter_effects} shows the $P(\nu_\mu \rightarrow \nu_e)$ appearance probabilities (left panel), $P(\nu_\mu \rightarrow \nu_\mu)$ survival probabilities (middle panel), and $P(\nu_\mu \rightarrow \nu_\tau)$ appearance probabilities (right panel) as a function of true neutrino energy $(E_\nu)$ and arrival direction ($\cos\theta_\nu$). The expected signal regions in the ($E_\nu$, $\cos\theta_\nu$) plane are identified by examining the variations in oscillation probabilities that arise when different Earth density models are compared with respect to 12-layered PREM model.

\begin{figure*}[htp!]
	\includegraphics[width=\linewidth]{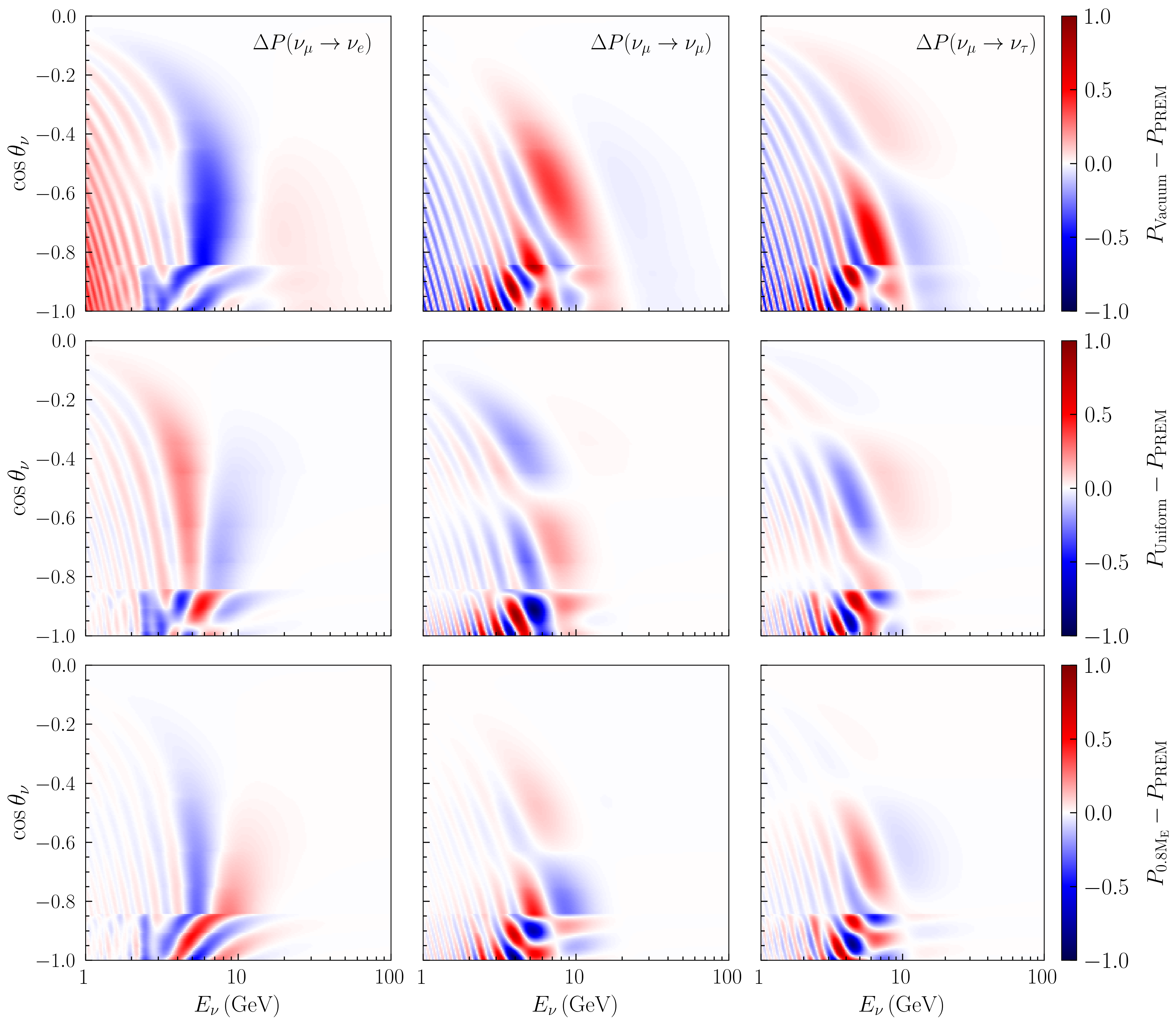}
	\caption{The difference of three-flavor oscillation probabilities between alternative Earth density models and the 12-layered PREM model, shown as a function of true neutrino energy ($E_\nu$) and arrival direction ($\cos\theta_\nu$), assuming normal mass ordering. The left, middle, and right columns correspond to the probability channels $P(\nu_\mu \rightarrow \nu_e)$, $P(\nu_\mu \rightarrow \nu_\mu)$, and $P(\nu_\mu \rightarrow \nu_\tau)$, respectively. The top row shows the differences between the vacuum oscillation scenario and the 12-layered PREM model. The middle row compares a uniform Earth density profile to the 12-layered PREM model. The bottom row shows the differences between a modified Earth density model with 20\% lighter mass ($0.8M_E$) and the 12-layered PREM model.}
	\label{fig:probability_diff}
\end{figure*}

In Fig.~\ref{fig:probability_diff}, the probability difference oscillograms illustrate how the signal regions are distributed across the ($E_\nu$, $\cos\theta_\nu$) parameter space for the different hypotheses under consideration: vacuum oscillations, a uniform density Earth, and an altered mass of the Earth. The first row shows the probability differences between vacuum oscillations and the 12-layered PREM model for the probability channels; $P(\nu_\mu \rightarrow \nu_e)$, $P(\nu_\mu \rightarrow \nu_\mu)$, and $P(\nu_\mu \rightarrow \nu_\tau)$. The second row illustrates the probability differences between a uniform Earth density model and the 12-layered PREM model, while the third row compares the oscillation probabilities between an Earth density profile with a 20\% lighter mass ($0.8 M_E$) and the 12-layered PREM model. In all the cases, the largest deviations occur at low energies and high baselines, where matter effects are most significant. These regions correspond to neutrinos traversing long paths through the Earth, particularly through the mantle and core, where the electron density modifies the oscillation probabilities. Therefore, the oscillations of multi-GeV neutrinos can probe the interior of the Earth.

\bibliography{References.bib}

\end{document}